\documentclass[12pt]{article}
\usepackage{amsmath}
\usepackage{graphicx}
\usepackage{natbib}
\usepackage[ruled,vlined]{algorithm2e}
\usepackage[colorlinks=false,hidelinks]{hyperref}
\usepackage{color}
\usepackage{setspace}
\usepackage{float}
\usepackage{rotating}
\usepackage{url} 
\usepackage[toc]{appendix}
\usepackage{minitoc}
\usepackage{adjustbox}
\usepackage{booktabs}
\newcommand{\blind}{0}

\begin{document}

\def\spacingset#1{\renewcommand{\baselinestretch}%
{#1}\small\normalsize} \spacingset{1}


\if0\blind
{
  \title{\bf Covariance Assisted Multivariate Penalized Additive Regression (CoMPAdRe)}
  \author{Neel Desai\\
    Division of Biostatistics, University of Pennsylvania\\
    and \\
    Veerabhadran Baladandayuthapani \\
    Department of Biostatistics, University of Michigan - Ann Arbor\\
    and\\
    Russell T. Shinohara \\ 
    Division of Biostatistics, University of Pennsylvania\\
    and\\
    Jeffrey S. Morris \\ 
     Division of Biostatistics, University of Pennsylvania\\}
  \maketitle
} \fi

\if1\blind
{
  \bigskip
  \bigskip
  \bigskip
  \begin{center}
    {\LARGE\bf Title}
\end{center}
  \medskip
} \fi
\vspace{-0.75cm}
\bigskip
\begin{abstract}
We propose a new method for the simultaneous selection and estimation of multivariate sparse additive models with correlated errors. Our method called Covariance Assisted Multivariate Penalized Additive Regression (CoMPAdRe) simultaneously selects among null, linear, and smooth non-linear effects for each predictor while incorporating joint estimation of the sparse residual structure among responses, with the motivation that accounting for inter-response correlation structure can lead to improved accuracy in variable selection and estimation efficiency.   
CoMPAdRe is constructed in a computationally efficient way that allows the selection and estimation of linear and non-linear covariates to be conducted in parallel across responses. Compared to single-response approaches that marginally select linear and non-linear covariate effects, we demonstrate 
in simulation studies that 
the joint multivariate modeling leads to gains in both 
estimation efficiency and selection accuracy, 
of greater magnitude in settings where signal is moderate relative to the level of noise. We apply our approach to protein-mRNA expression levels from  multiple breast cancer pathways obtained from The Cancer Proteome Atlas and characterize both mRNA-protein associations and protein-protein subnetworks for each pathway. We find non-linear mRNA-protein associations for the Core Reactive, EMT, PIK-AKT, and RTK pathways.  
\end{abstract}

\noindent%
{\it Keywords:} Multivariate analysis, Multivariate regression, Non-convex optimization, Variable selection, Semi-parametric regression 
\vfill

\spacingset{1.5} 
\section{Introduction}
\label{sec:intro}
Additive models are a generalization of linear models in which a response is modeled as the sum of arbitrary smooth, non-linear functions of covariates \citep{gammy, wood2017generalized}. Likewise, multivariate linear regression generalizes classical linear regression to the setting where $Q$ potentially correlated responses are regressed on a common set of $p$ predictors \citep{izenman2013multivariate}. Multivariate generalizations of the additive model are less common in the literature, and the setting of interest in this article. To set notation, let $\boldsymbol Y=\{\boldsymbol y_1, \ldots, \boldsymbol y_Q\}$ be a ($n \times Q$) matrix in which the $Q$ responses are potentially correlated and have a common set of $p$ predictors in ($n \times p$) matrix $\boldsymbol X=\{\boldsymbol X_1, \ldots, \boldsymbol X_p\}$. Here, we model 

\vspace{-.4in}
\begin{eqnarray}
\boldsymbol{y}_{q} = \beta_{q0} +  \sum_{j=1}^{p}f_{qj}(\boldsymbol X_{j})+ \boldsymbol e_{q} \hspace{0.2cm} \hbox{for} \hspace{0.2cm} q = 1,..., Q \label{eq:1}
\end{eqnarray}

\noindent where the $n$ rows of $\boldsymbol e=[\boldsymbol e_{1},...,\boldsymbol e_{Q}]$ are independently and identically $ N(\boldsymbol 0_{Q},\boldsymbol \Sigma_{Q \times Q}^{-1})$, and $\int f_{qj}(x) dx=0 \hspace{8pt} \forall (q,j)$ assumed for identifiability. Note that $f(\cdot)$ is indexed by response $q$ and covariate $j$. Just as in the classical single response setting of additive models, each response $\boldsymbol y_{q}$ is represented by the sum of smooth, covariate-specific functions $f_{qj}$ for each $\boldsymbol X_{j}$. In contrast to traditional settings, however, the residual error across the Q responses are correlated and related by precision matrix $\boldsymbol \Sigma_{Q \times Q}$. 


In the single response setting, a variety of approaches have been developed for variable selection in 
additive models. The majority of these methods have been $L_{1}$-based penalized regression procedures for selecting between null and non-linear predictor effects \citep{lin2006component, ravikumar2009sparse, huang2010variable}. Work has also been developed in a Bayesian context \citep{scheipl2012spike} and for models where estimated non-linear fits are piecewise constant with data-adaptive knots \citep{petersen2016fused}. More recently, research has been extended to include the selection between null, linear, and non-linear predictor effects involving variants of a group lasso penalty \citep{chouldechova2015generalized, lou2016sparse, petersen2019data}. These approaches often require pre-selection of a hyperparameter to favor linear versus non-linear fits. Multi-step algorithmic approaches have been developed as an alternative to favor parsimonious, interpretable linear fits when they sufficiently explain the association between predictor and response, with degree of preference towards linear fits controlled by a pre-specified hyperparameter \citep{tay2020reluctant}. 

There is also recent work on multivariate methods to perform simultaneous selection of regression coefficients and precision matrix elements to produce sparse solutions. For Gaussian multivariate linear regression, \cite{rothman2010sparse} and \cite{yin2013adjusting} proposed joint $L_{1}$ penalties on regression coefficients and off-diagonal elements of the precision matrix. In a Bayesian paradigm, \cite{bhadra2013joint}, \cite{ha2021bayesian} and \cite{consonni2017objective} proposed hierarchical models with a hyper-inverse Wishart prior on the covariance matrix, and 
\cite{deshpande2019simultaneous} 
used optimization based on a 
multivariate spike-and-slab lasso (mSSL) prior to simultaneously select sparse sets of linear coefficients and precision elements. 

There is some literature on methods for variable selection of non-linear functions in multivariate settings with non-independently distributed errors.
\cite{nandy2017additive} developed a method for the selection between null and smooth additive non-linear functions in the setting where errors are spatially dependent. The authors utilized the adaptive group lasso with an objective function that included a spatial weight matrix, establishing both selection consistency and convergence properties. In a multivariate context, \cite{niu2020bayesian} developed a  Bayesian variable selection procedure that, in the context of Gaussian graphical models, selected between null and smooth non-linear predictor effects for multiple responses while estimating the precision matrix among responses, provided that the precision matrix followed a decomposable graphical structure. This approach restricts all responses to have the same set of non-sparse predictors and does not distinguish between linear and non-linear predictor effects. To the best of our knowledge, the current literature does not have a multivariate approach that simultaneously selects among null, linear, and non-linear associations for each predictor-response pair while estimating an unconstrained dependence structure among responses.

To address this gap, we develop a computationally efficient solution to a sparse multivariate additive regression model, Covariance Assisted Multivariate Penalized Additive Regression (CoMPAdRe). CoMPAdRe employs a penalized spline basis representation  \citep{demmler1975oscillation} for $f_{qj}$, and jointly selects 
among null, linear, and smooth non-linear predictor effects for each response while simultaneously estimating a sparse precision matrix among responses. Our method enables the selection and estimation of linear and non-linear predictor effects to be conducted in parallel across responses. Through simulation studies,
we show that incorporating estimated residual structure into the selection and estimation of predictor effects leads to gains in selection accuracy and in statistical efficiency. We use CoMPAdRe to study the associations between mRNA and protein expression levels for 8 breast cancer pathways derived from The Cancer Proteome Atlas (TPCA) that reveal several non-linear associations missed by other approaches \citep{li2013tcpa}. Software for our method can be found at \url{https://github.com/nmd1994/ComPAdRe} along with examples for implementation.

\section{Methods}

To begin this section, we refer back to (1) and re-establish some notation. Let $\boldsymbol Y$ be a $(n \times Q)$ matrix with sample size n and Q possibly correlated responses. Similarly, let $\boldsymbol X$ be a ($n \times p$) matrix of $p$ covariates; we assume all covariates are continuous for exposition but $\boldsymbol X$ can also contain discrete covariates. Then, we model $\boldsymbol y_{q} = \beta_{q0} + \sum_{j=1}^{p}f_{qj}(\boldsymbol X_{j})+ \boldsymbol e_{q} \hspace{0.2cm} \hbox{for} \hspace{0.2cm} q = 1,..., Q$ where the $n$ rows of $\boldsymbol e=[\boldsymbol e_{1},...,\boldsymbol e_{Q}]$ are independently and identically $ N(\boldsymbol 0_{Q},\boldsymbol \Sigma_{Q \times Q}^{-1})$, and $\int f_{qj}(x) dx=0 \hspace{8pt} \forall (q,j)$ assumed for identifiability.  

Additive models can be written using 
a number of possible non-linear functional representations for each $f_{qj}(\cdot)$. Some common choices are Gaussian processes or smoothing splines \citep{cheng2019additive, wood2017generalized}. Our multivariate framework is based on using the Demmler Reinsch parameterization of smoothing splines or O’Sullivan penalized splines, which naturally partitions 
linear from non-linear components. In the next sub-sections, we introduce our notation and formulation starting with the notation of spline representation, the univariate additive model, and our variable selection processes before fully specifying our multivariate additive model selection approach. 


\subsection{Additive Models with Penalized Splines}
 A smoothing spline is a bivariate smoother for a single ($n \times 1$) response and covariate \{\textbf{y}, $\boldsymbol X_{j}$\} whose objective is described as follows: $\hbox{min} \hspace{0.1cm} L(f_{j})=\{\boldsymbol y - f_{j}(\boldsymbol X_{j})\}^{T}\{\boldsymbol y - f_{j}(\boldsymbol X_{j})\} + \lambda_{j} || f_{j}^{''}||_{2}^{2}$. Smoothness is determined by tuning parameter $\lambda_{j}$, which controls the balance between interpolation and having a small second derivative; $f_{j}$ is a linear function when $||f_{j}^{''}||_{2}^{2}=0$ \citep{wang2011smoothing}. The optimal solution to the smoothing spline problem can be obtained by solving a representation of $f_{j}$ determined by a ($n \times k$) basis of splines $\boldsymbol \Phi_{j} = (\boldsymbol\phi_{1j}|\boldsymbol \phi_{2j}|...|\boldsymbol \phi_{kj})$ with $k$ knots defined at unique values of $\boldsymbol X_{j}$: $f_{j} = \sum_{k=1}^{n}\boldsymbol\phi_{kj}\beta_{kj}$. To further establish notation, let $\boldsymbol \beta_{j} = (\beta_{1j}|\beta_{2j}|...|\beta_{kj})$ be a ($k \times 1$) vector of coefficients, then $\hat{f_{j}} = \boldsymbol \Phi_{j} \hat{\boldsymbol \beta_{j}}$. Our objective is now  min $L(f_{j})=\{\boldsymbol y - f_{j}(\boldsymbol X_{j})\}^{T}\{\boldsymbol y - f_{j} (\boldsymbol X_{j})\} + \lambda_{j}||f_{j}^{''}||_{2}^{2}= (\boldsymbol y- \boldsymbol\Phi_{j} \boldsymbol \beta_{j})^{T}(\boldsymbol y - \boldsymbol\Phi_{j} \boldsymbol \beta_{j})+\lambda_{j}\boldsymbol \beta_{j}^{T} \boldsymbol \Omega_{j} \boldsymbol \beta_{j}$ where $\Omega_{ij}= \int  \boldsymbol \phi(z)_{i}^{''} \boldsymbol \phi(z)_{j}^{''}dz$. This is a standard penalized least squares problem, whose solution is $\hat{\boldsymbol \beta_{j}}=(\boldsymbol \Phi_{j}^{T} \boldsymbol \Phi_{j} + \lambda_{j}\boldsymbol \Omega_{j})^{-1}\boldsymbol \Phi_{j}^{T} \boldsymbol y$ \citep{wang2011smoothing, Hansen2019}. For computational efficiency one can use ($k<n$) knots defined at certain quantiles of $\boldsymbol X_{j}$, which would then be called penalized splines. We utilize O'Sullivan penalized splines for this article \citep{wand2008semiparametric}.
 
 Given the detailed description above, we define a penalized objective for an additive model where each function $f_{j}$ has a spline representation. We see that

\[ \hbox{\textbf{min}} \hspace{0.2cm} \{\boldsymbol y - \beta_{0} - \sum_{j=1}^{p}f_{j}(\boldsymbol X_{j})\}^{T}\{\boldsymbol y - \beta_{0} -\sum_{j=1}^{p}f_{j}(\boldsymbol X_{j})\} + \sum_{j=1}^{p}\lambda_{j}\int f_{j}^{''}(t_{j})^{2}dt_{j} \] 

\noindent Conveniently, the optimal minimizer to this problem is an additive cubic spline model for which each $f_{j}$ is a cubic spline with knots defined at unique values of each $\boldsymbol X_{j}$ \citep{wang2011smoothing,wood2017generalized}. Note that we assume $\sum_{i=1}^{n}f_{j}(x_{ij})=0$ for all $p$ in order to ensure an identifiable solution. To represent each $f_{j}$, we next outline the Demmler-Reinsch representation before redefining our additive model objective. 


 To form the Demmler-Reinsch representation of a smoothing (or penalized) spline, which separates linear and non-linear components of function $f_{j}$, we revisit spline basis $\boldsymbol \Phi_{j}$. Specifically, we note that $\hbox{svd}(\boldsymbol\Phi_{j})=\boldsymbol U_{j} \boldsymbol D_{j}\boldsymbol V_{j}^{T}$, where $\boldsymbol U_{j}$ is a ($n \times k$) orthogonal matrix, $\boldsymbol D_{j}$ is a diagonal ($k \times k$) matrix with all non-negative values, and $\boldsymbol V_{j}$ is a ($k \times k$) orthogonal matrix. We then rewrite $\hat{f_{j}}=\boldsymbol\Phi_{j}\hat{\boldsymbol \beta_{j}}=\boldsymbol U_{j} \boldsymbol D_{j}( \boldsymbol D_{j}^{2}+\lambda_{j} \boldsymbol V_{j}^{T} \boldsymbol \Omega_{j} \boldsymbol V_{j})^{-1}\boldsymbol D_{j} \boldsymbol U_{j}^{T}= \boldsymbol U_{j}(\boldsymbol I - \lambda_{j} \widetilde{\boldsymbol \Omega_{j}})^{-1}\boldsymbol U_{j}^{T}$. Note that $\widetilde{\boldsymbol \Omega_{j}}=\boldsymbol D_{j}^{-1}\boldsymbol V_{j}^{T} \boldsymbol \Omega_{j} \boldsymbol V_{j} \boldsymbol D_{j}^{-1}$ is a ($k \times k$) positive semi-definite matrix that can be diagonalized as $\hbox{svd}(\widetilde{\boldsymbol \Omega_{j}})=\boldsymbol W_{j} \boldsymbol \Gamma_{j} \boldsymbol W_{j}^{T}$. $\boldsymbol W_{j}$ is a ($k \times k$) orthogonal matrix and $\boldsymbol \Gamma_{j}$ is a ($k \times k$) diagonal matrix with decreasing non-negative elements $(\gamma_{1}|\gamma_{2}|...|0|0)$ whose last two diagonal entries are zero \citep{wood2017generalized, Hansen2019}. Given these representations, 
we now re-write $\hat{f_{j}}=\boldsymbol U_{j} \boldsymbol W_{j}(\boldsymbol I+\lambda_{j} \boldsymbol \Gamma_{j})^{-1}\boldsymbol W_{j}^{T} \boldsymbol U_{j}^{T}=\widetilde{\boldsymbol U_{j}}(\boldsymbol I + \lambda_{j} \boldsymbol \Gamma_{j})^{-1}\widetilde{\boldsymbol U_{j}}^{T}$. $\widetilde{\boldsymbol U_{j}}$ is the ($n \times k$) Demmler-Reinsch basis with corresponding ($k \times 1$) coefficient vector $\Tilde{\boldsymbol \beta_{j}}$. The final two columns of $\widetilde{\boldsymbol U_{j}}$ correspond to the intercept and linear components of $f_{j}$, and the other columns of $\widetilde{\boldsymbol U_{j}}$ correspond to non-linear components of $f_{j}$. The now diagonal smoothing penalty $\boldsymbol \Gamma_{j}$ leaves the 2 intercept and linear basis functions of $\widetilde{\boldsymbol U_{j}}$ unpenalized and the remaining basis functions increasingly penalized by their complexity (i.e. the values along the diagonal of $\boldsymbol \Gamma_{j}$) \citep{demmler1975oscillation, Hansen2019}. 



To summarize, we can represent the objective for $f_{j}$ as follows. Let $\widetilde{\boldsymbol U_{j}}= [\boldsymbol 1, \Tilde{\boldsymbol U_{j}}^{L}, \hspace{0.1cm} \Tilde{\boldsymbol U_{j}}^{NL}]$ denote the linear and non-linear components of basis functions $\widetilde{\boldsymbol U_{j}}$, respectively, with corresponding coefficients $\widetilde{\boldsymbol\beta_{j}}=[\beta_{0}, \hspace{0.1cm} \Tilde{ \beta_{j}}^{L},\hspace{0.1cm} \Tilde{\boldsymbol\beta_{j}}^{NL}]$. Since $\boldsymbol X_{j}$ is just an orthogonal rotation of $\Tilde{\boldsymbol U_{j}}^{L}$, we can trivially replace $\Tilde{\boldsymbol U_{j}}^{L}$ with $\boldsymbol X_{j}$ and see:
\begin{center}{\textbf{minimize} $\{\boldsymbol y-( \beta_{0} + \boldsymbol X_{j}\Tilde{ \beta_{j}}^{L} + \Tilde{\boldsymbol U_{j}}^{NL}\Tilde{\boldsymbol \beta_{j}}^{NL})\}^{T}\{\boldsymbol y -(\beta_{0} + \boldsymbol X_{j} \Tilde{ \beta_{j}}^{L} + \Tilde{\boldsymbol U_{j}}^{NL}\Tilde{\boldsymbol \beta_{j}}^{NL})\} + \lambda_{j} \Tilde{\boldsymbol\beta_{j}^{T}} \boldsymbol \Gamma_{j}\Tilde{\boldsymbol \beta_{j}}$}
\end{center}

\noindent and can then rewrite our additive model objective with $j = 1,...,p$ covariates as:

\begin{center}{\textbf{minimize} \small$\{\boldsymbol y - \beta_{0} - \sum_{i=j}^{p} (\boldsymbol X_{j} \Tilde{\beta_{j}}^{L} + \Tilde{\boldsymbol U_{j}}^{NL}\Tilde{\boldsymbol \beta_{j}}^{NL})\}^{T}\{\boldsymbol y - \beta_{0} - \sum_{j=1}^{p}(\boldsymbol{X_{j}}\Tilde{\beta_{j}}^{L} + \Tilde{\boldsymbol U_{j}}^{NL}\Tilde{\boldsymbol\beta_{j}}^{NL})\} + \sum_{i=j}^{p}\lambda_{j}\Tilde{\boldsymbol \beta_{j}^{T}}\boldsymbol\Gamma_{j}\Tilde{\boldsymbol\beta_{j}}$}
\end{center}

\noindent To ensure identifiability constraints are satisfied, we omit all intercept basis functions for each $f_{j}$ and fit one global intercept \citep{lee2018bayesian}. We now conclude this sub-section by briefly describing how we can fit additive models as a linear mixed model; this detail becomes relevant during the estimation portion of our framework.

\paragraph{Estimating Additive Models as a Linear Mixed Model:}  One can represent any $f_{j}=\sum_{k=1}^{n}\boldsymbol \phi_{kj}(x) \boldsymbol \beta_{kj}$ with an associated measure controlling `wiggliness' (i.e. $\boldsymbol\beta_{j}^{T}\boldsymbol \Omega_{j} \boldsymbol \beta_{j}$) in terms of a mixed effects model if they can separate linear and non-linear components of $f_{j}$, treating the linear components as fixed effects and non-linear components as random effects \citep{wood2017generalized}. Given the Demmler-Reinsch representation, we can then represent a smoothing spline or penalized spline as follows:  $\boldsymbol y = \boldsymbol X_{j}\Tilde{ \beta_{j}}^{L} + \Tilde{\boldsymbol U_{j}}\Tilde{\boldsymbol \beta_{j}}^{NL}+\boldsymbol \epsilon$, where $\Tilde{\boldsymbol \beta_{j}}^{NL} \sim N(\boldsymbol 0,\frac{\boldsymbol \Gamma_{NL}^{-1}}{ \lambda_{j}})$.  $\boldsymbol \Gamma_{NL}$ represents a diagonal matrix of the positive non-zero values of $\boldsymbol \Gamma$, and $\epsilon \sim N(\boldsymbol 0, \boldsymbol I_{n} \sigma^{2})$. To incorporate an additive model into the mixed model framework, one can append the linear components for each function $f_{j}$ to the matrix of fixed effects and non-linear components of $f_{j}$ to the random effects design matrix while ensuring identifiability by restricting to a global intercept \citep{wood2017generalized}.

\subsection{Variable Selection for Additive Models}

Given that additive models consider multiple covariates whose dimension $p$ may be high, approaches that perform variable selection can greatly aid in producing interpretable, computationally feasible solutions. 
To perform variable selection for functions $f_{j}$, in this article we adapt a smoothness-sparsity penalty that uses separate hyperparameters to control smoothness and selection of additive models. Adopting the above notation, the general form of the penalty is as follows: $||\boldsymbol y-\sum_{j=1}^{p}f_{j}||_{n}^{2} + \sum_{j=1}^{p}J(f_{j})$, where $||f_{j}||_{n}^{2}=\frac{1}{n}\sum_{i=1}^{n}f_{ij}^{2}$ and $J(f_{j})=\lambda_{1}\sqrt{||f_{j}||_{n}^{2}+\lambda_{2}\int (f_{j}^{''}(x))^{2}dx}$. Penalty $J(f_{j})$ has two hyperparameters: $\lambda_{1}$ controls the level of sparsity while $\lambda_{2}$ controls smoothness via the same second-derivative penalty utilized in smoothing splines. Using the spline basis formulation for each $f_{j}$ previously described, the method's objective is
\textbf{min} $||\boldsymbol y- \beta_{0} - \sum_{j=1}^{p}\boldsymbol \Phi_{j}\boldsymbol \beta_{j}||_{n}^{2}+\lambda_{1}\sum_{j=1}^{p}\sqrt{\frac{1}{n}\boldsymbol \beta_{j}^{T}\boldsymbol \Phi_{j}^{T}\boldsymbol \Phi_{j}\boldsymbol \beta_{j}+ \lambda_{2}\boldsymbol \beta_{j}^{T}\boldsymbol \Omega_{j}\boldsymbol \beta_{j}}$. Computationally, this objective can be re-parameterized as a simple group lasso problem for fixed $\lambda_{2}$, ensuring the smoothness of the functional fit is accounted for in the selection process \citep{meier2009high}.

To illustrate how we perform linear and non-linear predictor selection, we first present a single response objective in this section before moving on to our full multivariate objective, where we induce sparsity in the precision matrix with $L_{1}$ penalization.  More concretely, we utilize the Demmler-Reisch basis to separate linear and non-linear components of $f_{j}$, penalizing linear basis functions with an $L_{1}$ penalty and non-linear basis functions with an adapted version of the smoothness-sparsity penalty outlined above \citep{meier2009high}. We now note the following objective:  \begin{center}{\hspace{4cm}\textbf{minimize} \newline $||\boldsymbol y - \beta_{0} - \sum_{j=1}^{p}(\boldsymbol X_{j}\Tilde{ \beta_{j}}^{L}+\Tilde{\boldsymbol U_{j}}^{NL}\Tilde{\boldsymbol\beta_{j}}^{NL})||_{2}^{2}+ \lambda_{1}\sum_{j=1}^{p}||\Tilde{ \beta_{j}}^{L}||_{1} \newline \newline + \lambda_{2} \sum_{j=1}^{p} \sqrt{\frac{1}{n}(\Tilde{\boldsymbol\beta_{j}}^{NL})^{T}(\Tilde{\boldsymbol U_{j}}^{NL})^{T}\hspace{0.1cm}\Tilde{\boldsymbol U_{j}}^{NL}\Tilde{\boldsymbol\beta_{j}}^{NL}+ \lambda_{3j}(\Tilde{\boldsymbol \beta_{j}}^{NL})^{T}\hspace{0.1cm}\boldsymbol \Gamma_{NL}^{j}\hspace{0.1cm}\Tilde{\boldsymbol \beta_{j}}^{NL}}$}
 \end{center}

Linear selection is controlled by hyperparameter $\lambda_{1}$ while non-linear selection is controlled by $\lambda_{2}$. hyperparameter $\lambda_{3j}$ controls the degree of smoothness for covariate $j$'s non-linear basis functions; note that we use $\Gamma_{NL}^{j}$ because the first two diagonal components of $\Gamma$, corresponding to linear basis functions, are zero (meaning linear components are left unsmoothed). 
This optimization problem will produce a selection of null, linear, or nonlinear effect for each proposed covariate $j=1,\ldots,p$, and in the univariate additive model settings we refer to this approach as \textit{Penalized Additive Regression}, or \textbf{PAdRe}.  In simulation studies, to show the benefit of multivariate modeling we will use this as a comparitor for out multivariate approach {\bf CoMPAdRe}, which we will now describe.


\subsection{Multivariate Objective and Model Fitting Procedure}
Now transitioning to the multivariate additive model setting, suppose we observe a ($n \times Q$) matrix $\boldsymbol Y$ where each of our $Q$ responses are potentially correlated and contain a common ($n \times p$) predictor set $\boldsymbol X$. We begin by identifying a joint likelihood for our model. Let $\boldsymbol y_{q} = \sum_{j=1}^{p}f(\boldsymbol X_{qj})+ \boldsymbol e_{q}$ for $q = 1,..., Q$ where
the $n$ rows of $\boldsymbol e=[\boldsymbol e_{1},...,\boldsymbol e_{Q}]$ are independently and identically $ N(\boldsymbol 0_{Q},\boldsymbol \Sigma_{Q \times Q}^{-1})$. Then, we can rewrite $\boldsymbol y_{q} = \beta_{0} + \sum_{j=1}^{p} \boldsymbol X_{j} \Tilde{\beta_{qj}}^{L} + \Tilde{\boldsymbol U_{j}}^{NL}\Tilde{\boldsymbol\beta_{qj}}^{NL}+ \boldsymbol e_{q}$. Extending to Q responses, we concatenate notation and redefine $\boldsymbol X = [\boldsymbol 1,\hspace{0.1cm} \boldsymbol X], \hspace{0.2cm} \Tilde{\boldsymbol \beta}^{L} = [\boldsymbol 1,\hspace{0.1cm}\Tilde{\boldsymbol \beta}^{L}]$ and see:
\vspace{-.4in}
\begin{center}
$\boldsymbol Y \sim MVN(\boldsymbol X \Tilde{\boldsymbol\beta}^{L} + \Tilde{\boldsymbol U}^{NL}\Tilde{\boldsymbol \beta}^{NL},\hspace{0.1cm}\boldsymbol \Sigma_{Q\times Q}^{-1})$
\end{center}
\noindent with the following negative log-likelihood:

\begin{center}
$g(\Tilde{\boldsymbol \beta}, \boldsymbol \Sigma)=tr[\frac{1}{n}\{\boldsymbol Y - (\boldsymbol X\Tilde{\boldsymbol \beta}^{L} + \Tilde{\boldsymbol U}^{NL}\Tilde{\boldsymbol\beta}^{NL})\}^{T}\{\boldsymbol Y - (\boldsymbol X\Tilde{\boldsymbol \beta}^{L} + \Tilde{\boldsymbol U}^{NL}\Tilde{\boldsymbol \beta}^{NL})\}\boldsymbol\Sigma]- log |\boldsymbol \Sigma|$
\end{center}
\noindent where $\Tilde{\boldsymbol U}^{NL}$ is a concatenated [$n \times p(k-2)$] matrix of non-linear basis functions, where each of p covariates has $k-2$ non-linear basis functions. 

We now introduce penalties to our negative log-likelihood, completing our objective. Along with the penalties outlined in the previous section, we include an $L_{1}$ penalty on the off-diagonal elements of precision matrix $\boldsymbol \Sigma_{Q\times Q}$, similar to those utilized in the graphical lasso \citep{friedman2008sparse}. Our final model objective is demonstrated as follows: 

\begin{center}{\hspace{4cm}\textbf{minimize} \newline $g(\Tilde{\boldsymbol \beta},\boldsymbol \Sigma) + \sum_{q=1}^{Q}\lambda_{1q}\sum_{j=1}^{p}||\Tilde{\boldsymbol \beta}_{qj}^{L}||_{1} +  \newline \newline \sum_{q=1}^{Q}\lambda_{2q} \sum_{j=1}^{p} \sqrt{\frac{1}{n}(\Tilde{\boldsymbol\beta}_{qj}^{NL})^{T}(\Tilde{\boldsymbol U}_{qj}^{NL})^{T}\hspace{0.1cm}\Tilde{\boldsymbol U}_{qj}^{NL}\Tilde{\boldsymbol \beta}_{qj}^{NL}+ \lambda_{3qj}(\Tilde{\boldsymbol \beta}_{qj}^{NL})^{T}\hspace{0.1cm} \boldsymbol\Gamma_{NL}^{qj}\hspace{0.1cm} \Tilde{\boldsymbol\beta}_{qj}^{NL})} \hspace{3.2cm} (2) \newline \newline +\lambda_{4}\sum_{q^{'}} ||\sigma_{q^{'}, q}||_{1}$}
\end{center}
\vspace{1cm}

To produce a solution to this non-convex objective, we break our problem down into a three step procedure in which each step, conditioned on the other steps, reduces to a simpler convex optimization problem. Our model fitting procedure also enables parallel computing in terms of responses $q$, ensuring a computationally scalable solution. We next describe this procedure, and how model-fitting is implemented, in more detail. 

\subsection{Model Fitting Procedure}
Before outlining model fitting in \textbf{Algorithm 1}, we first describe a parameterization of precision matrix $\boldsymbol \Sigma_{Q\times Q}$ using a formulation that 
is based on the relationship between the precision matrix of a multivariate normal distribution and regression coefficients \citep{anderson1984introduction}. Specifically, let $\{n \times (Q-1)\}$ matrix $\boldsymbol E_{q'}$ denote the 
vector of errors $[\boldsymbol{e_{1},\ldots,e_{q-1},e_{q+1},\ldots ,e_{Q}}]^T$ other than $\boldsymbol e_{q}$, the error term for response $q$. We can then represent $\boldsymbol e_{q}$ in terms of the following regression: $\boldsymbol e_{q}=\boldsymbol E_{q'} \boldsymbol \alpha_{q} + \boldsymbol \epsilon_{q}$, where coefficient $\boldsymbol \alpha_{q}= - \frac{\boldsymbol \Sigma_{q, \boldsymbol E_{q'}}}{\sigma_{q,q}}$, $\boldsymbol \Sigma_{q, \boldsymbol E_{q'}}$ is a $((Q-1) \times 1)$ vector of partial correlations of response $q$ with the other $Q-1$ responses, and error term $\boldsymbol \epsilon_{q} \sim N(0, \frac{1}{\sigma_{q,q}})$. Recent literature has extended this concept to the topic of variable selection for multi-layered Gaussian graphical models (mlGGM) \citep{ha2021bayesian}. Their work, in the context of a multivariate regression, 
demonstrated that model fitting could be conducted as parallel single response regressions. 

Specifically, 
in our context we can rewrite each response of $\boldsymbol Y$ as follows: 

\begin{equation} \tag{3} \boldsymbol Y_{q} = (\boldsymbol X \Tilde{\boldsymbol \beta}_{[,q]}^{L}+ \Tilde{\boldsymbol U}^{NL}\Tilde{\boldsymbol \beta}_{[,q]}^{NL})+ \boldsymbol Y_{[,-q]} \boldsymbol \alpha_{q} - \boldsymbol X\Tilde{\boldsymbol \beta}_{[,-q]}^{L}\boldsymbol \alpha_{q} - \Tilde{\boldsymbol U}^{NL}\Tilde{\boldsymbol \beta}_{[,-q]}^{NL}\boldsymbol \alpha_{q} + \boldsymbol \epsilon_{q}\end{equation}

\noindent where $[,q]$ denotes the column corresponding to response $q$ and $[,-q]$ refers to the set of all columns other than the column indexing response $q$. Given estimates of linear coefficients, non-linear coefficients, and precision matrix $\boldsymbol \Sigma_{Q \times Q}$, this formulation enables updating linear and non-linear selection and estimation steps as parallel single-response procedures, and
reduces the selection of linear, non-linear, and precision components to simpler convex optimization problems when conditioned on previous estimates.

More specifically, \textbf{Algorithm 1} breaks our overall model objective into 3 sequential steps given initial estimates of $\Tilde{\boldsymbol\beta}^{L}$, $\Tilde{\boldsymbol\beta}^{NL}$, and $\boldsymbol \Sigma$. Rewriting each response of $\boldsymbol Y$ as seen above in (3), we first condition on ($\Tilde{\boldsymbol\beta}^{NL}, \boldsymbol \Sigma$) and update $\Tilde{\boldsymbol\beta}^{L}$ via $Q$ separate Lasso procedures; selected coefficients are then re-estimated via OLS. Then, conditioning on ($\Tilde{\boldsymbol\beta}^{L},\boldsymbol \Sigma$) we update selection of $\Tilde{\boldsymbol\beta}^{NL}$ via $Q$ separate group lasso procedures; both $\Tilde{\boldsymbol\beta}^{L}$ and $\Tilde{\boldsymbol\beta}^{NL}$ are then re-estimated with $Q$ separate linear mixed models. Finally, conditioning on ($\Tilde{\boldsymbol\beta}^{L}, \Tilde{\boldsymbol\beta}^{NL}$) we update $\Sigma$ with the graphical lasso. We iterate between these three steps until $\hbox{MSE}(\hat{Y})$ converges within a pre-specified tolerance. All tuning parameters are selected via cross-validation; full details on how cross-validation is implemented can be seen at the end of supplemental section \textbf{S.1}. Smoothness hyperparameters $\lambda_{3qj}$ are pre-specified with generalized cross-validation (GCV) marginally for every covariate-response combination; pre-fixing every $\lambda_{3qj}$ enables the selection of $\Tilde{\boldsymbol\beta}^{NL}$ to be reduced to $Q$ separate group lasso problems when conditioned on $\Tilde{\boldsymbol\beta}^{L}$ and $\boldsymbol \Sigma$. We note that CoMPAdRe does not pre-specify a preference for linear vs. non-linear fits; both linear and non-linear selection is performed for every covariate-response combination. 


\begin{algorithm}[H]
\small
\setstretch{1.4}
\SetAlgoLined
\KwResult{Mean Components: Linear: $\Tilde{\boldsymbol \beta}^{L}$, Non-Linear: $\Tilde{\boldsymbol \beta}^{NL}$ $\&$ Estimated Precision $\boldsymbol \Sigma$}
\textbf{Require} Initial estimates $\Tilde{\boldsymbol\beta}^{L}$, $\Tilde{\boldsymbol\beta}^{NL}$, and $\boldsymbol \Sigma$ \;
\For{q = 1,...,Q}{
  Set $\boldsymbol Y_{q}^{*}=\boldsymbol Y_{q} - \Tilde{\boldsymbol U}^{NL}\Tilde{\boldsymbol \beta}_{[,q]}^{NL} - \boldsymbol Y_{[,-q]}\boldsymbol \alpha_{q} + \boldsymbol X\Tilde{\boldsymbol \beta}_{[,-q]}^{L}\boldsymbol \alpha_{q} + \Tilde{\boldsymbol U}^{NL}\Tilde{\boldsymbol \beta}_{[,-q]}^{NL}\boldsymbol\alpha_{q}$  \;
 Solve $min_{\boldsymbol \beta_{[,q]}^{L}} \hspace{0.2cm} \|\boldsymbol Y_{q}^{*} - \boldsymbol X \Tilde{\boldsymbol\beta}_{[,q]}^{L} \| + \lambda_{1q}||\Tilde{\boldsymbol \beta}_{[,q]}^{L}||_{1} $ \;
 }
Step 1.5: Linear Re-estimation with OLS \;
Let $\boldsymbol X_{[,q]}^{\hbox{select}}$, $\boldsymbol\beta_{[,q]}^{\hbox{select}}$ denote the subset of covariates and coefficients selected as non-zero for response q \;
\For{q = 1,...,Q}{
$min_{\boldsymbol \beta_{[,q]}} \hspace{0.2cm}\|\boldsymbol Y_{q}^{*} - \boldsymbol X_{[,q]}^{\hbox{select}} \boldsymbol \beta_{[,q]}^{\hbox{select}}\|$
}
 Step 2: Non-Linear Selection Update\; 
 \For{q = 1,...,Q}{
  Set $\boldsymbol Y_{q}^{*}=\boldsymbol Y_{q}- \boldsymbol X\Tilde{\boldsymbol\beta}_{[,q]}^{L} - \boldsymbol Y_{[,-q]}\boldsymbol \alpha_{q} + \boldsymbol X \Tilde{\boldsymbol \beta}_{[,-q]}^{L}\boldsymbol \alpha_{q} + \Tilde{\boldsymbol U}^{NL}\Tilde{\boldsymbol \beta}_{[,-q]}^{NL} \boldsymbol\alpha_{q}$  \;
 Solve $min_{\Tilde{\boldsymbol \beta}_{[,q]}^{NL}} \hspace{0.2cm} \|\boldsymbol Y_{q}^{*} - \Tilde{\boldsymbol U}^{NL}\Tilde{\boldsymbol \beta}_{[,q]}^{NL}\| + \lambda_{2q}\sum_{j=1}^{p}\sqrt{\frac{1}{n}(\Tilde{\boldsymbol \beta}_{[j,q]}^{NL})^{T}(\Tilde{\boldsymbol U}_{qj}^{NL})^{T}\Tilde{\boldsymbol U}_{qj}^{NL}\Tilde{\boldsymbol\beta}_{[j,q]}^{NL} + \lambda_{3qj}(\boldsymbol \beta_{[j,q]}^{NL})^{T}\boldsymbol\Gamma_{NL}^{qj}\boldsymbol \beta_{[j,q]}^{NL}} $ \;
 }

 Step 2.5: Linear and Non-Linear Coefficient Estimation with Mixed Effects Model\;
 \For{q = 1,...,Q}{
Set $\boldsymbol Y_{q}^{*}= \boldsymbol Y_{q} - \boldsymbol Y_{[,-q]}\boldsymbol \alpha_{q} + \boldsymbol X \Tilde{\boldsymbol \beta}_{[,-q]}^{L}\boldsymbol \alpha_{q} + \Tilde{\boldsymbol U}^{NL}\Tilde{\boldsymbol \beta}_{[,-q]}^{NL} \boldsymbol\alpha_{q}$\;

Let $\boldsymbol U_{[,q]}^{\hbox{n-select}}$, $\boldsymbol \beta_{[,q]}^{\hbox{n-select}}$ denote the subset of basis functions and coefficients selected as nonzero and non-linear for response q \;  

Update estimation of $\boldsymbol \beta_{[,q]}^{\hbox{select}}$, $\boldsymbol \beta_{[,q]}^{\hbox{n-select}}$ with a linear mixed model
 }
 Step 3: Selection of Precision Elements\;
 Let $\boldsymbol Y^{*} = \boldsymbol Y - X\Tilde{\boldsymbol \beta}^{L} - \Tilde{\boldsymbol U}^{NL} \Tilde{\boldsymbol \beta}^{NL}$ \;
Obtain an estimate of the empirical covariance and update precision matrix $\boldsymbol \Sigma$ using the graphical lasso \citep{friedman2008sparse}\;
 
 Stop when $\hbox{MSE}_{\boldsymbol Y}$ converges within \textbf{tol}
 
 \caption{Covariance-Assisted Multivariate Sparse Additive Regression (CoMPAdRe)}
\end{algorithm}


\section{Simulation Study}
\paragraph{Simulation Design:} We assess CoMPAdRe's performance under settings with varying sample sizes, levels of residual dependence, and signal-to-noise ratios. We consider sample size $n = 250$, number of responses $Q = 10$, fix the number of covariates to either $p=10$ or $p=100$, and use the following signal-to-noise ratios: $\delta = (0.25, 0.5, 0.75, 1, 2)$. To induce residual dependence, we specify a Toeplitz structure to generate covariance matrices (in this case correlation matrices): $\Sigma^{-1}=(\rho^{|k-k'|})_{k,k'=1}^{q}$, where $\rho = (0.2, 0.5, 0.7, 0.9)$ and higher values of $\rho$ correspond to more highly dependent responses. We present results for $p = 100$ in the main body of this manuscript and present results from all other settings in supplemental section \textbf{S.1}. Each covariate $\boldsymbol{X_{j}}$, for $j = 1, ..., p$, is generated using $n$ draws from a random uniform distribution: $\hbox{Unif}(-1,1)$. We consider the following functions for non-null associations between a covariate and a response, four types of nonlinear functions: $\boldsymbol{f_{1}}=\delta*(1-\hbox{exp}(-2X_{j}))$,  $\boldsymbol{f_{2}}=\delta*(X_{j}^{2})$, $\boldsymbol{f_{3}}= \delta*(X_{j}^{3})$, $\boldsymbol{f_{4}}(\sigma = 0.1)= \delta*\frac{1}{\sqrt{2\pi}\sigma}\hbox{exp}(-\frac{X_{j}^{2}}{2\sigma^{2}}))$, 
as well as linear functions: $\boldsymbol{f_{5}}= \delta*(X_{j})$. A figure displaying the shapes of all functions along with additional simulations assessing function-specific selection and estimation performance across all methods considered is contained in supplemental section \textbf{S.1.2}. 



For each simulated dataset, we generate the sparse set of predictors for each response as follows. We first randomly select 4 of the first 5 responses to have non-zero predictors. For each non-sparse response, we then randomly select 1 - 5 covariates to have any signal with that response. Each of these selected covariates is randomly assigned a function with probability $0.125$ for $f_1, \ldots, f_4$ and $0.5$ for $f_5$. Each response is then be generated as $Y_{iq}=\sum_{j=1}^{p} f_{jq}(X_{ij}) + E_{iq}; i=1,\ldots,n; q=1,\ldots,Q$, where $\boldsymbol{E_i}=(E_{i1}, \ldots E_{iQ})^T \sim MVN(\boldsymbol{0_{Q}},\Sigma^{-1})$. For each setting of sample size $n$, residual dependence level $\rho$, and signal-to-noise $\delta$ considered, we simulate 50 datasets in the manner outlined above.

\paragraph{Performance Assessment:} We compare CoMPAdRe to the following approaches: \textbf{(1)} PAdRe, \textbf{(2)} GAMSEL \citep{chouldechova2015generalized}, \textbf{(3)} Lasso \citep{tibshirani1996regression}, and \textbf{(4)} mSSL \citep{deshpande2019simultaneous}. These approaches can be divided as follows: single response approaches that marginally select linear and non-linear covariate associations for
each response (\textbf{1, 2}), single response approaches that marginally select linear covariate associations ignoring the distinction between linear and non-linear fits (\textbf{3}), and multivariate approaches that simultaneously select linear associations and precision elements  ignoring the distinction between linear and non-linear fits (\textbf{4}). Note that GAMSEL has a user-selected parameter for favoring linear vs. non-linear fits - we used the approach's suggested default. All other tuning parameters across methods were selected via cross-validation (further described in supplemental section \textbf{S.1}). For \textbf{selection accuracy}, we report the true positive rate (TPR) and false positive rate (FPR) for null vs non-null signal. For \textbf{estimation accuracy}, we present the ratio of mean absolute deviation (MAD) between CoMPAdRe and method \textbf{(1)}, $(\footnotesize{\frac{\hbox{CoMPAdRe}}{\hbox{PAdRe}}})$, for estimating true function $\Hat{f}$. Ratios of estimation accuracy between CoMPAdRe and other methods considered can be seen in supplemental section \textbf{S.1}. After selection the marginal approach \textbf{(1)} is re-estimated using a linear mixed effects model as in CoMPAdRe. 

\paragraph{Simulation Results:}

Table 1 summarizes our key selection results, reporting the true positive rate (TPR) and false positive rate (FPR) for all five methods at different levels of residual dependence $\rho$ and signal-to-noise ratio $\delta$. A comprehensive breakdown of selection results by type of function (linear vs. non-linear) can be seen in supplemental section \textbf{S.1}. Similarly, Figure 1 visualizes key estimation accuracy results across levels of $\rho$ and $\delta$ for estimating overall true signal $\Hat{f}$.

At moderate to high levels of residual dependence $(\boldsymbol{\rho = 0.7, 0.9})$, CoMPAdRe consistently demonstrates superior sensitivity for selecting significant covariates than other approaches while maintaining favorable false positive rates. The relative improvement of CoMPAdRe over PAdRe, an equivalent method in every way except ignoring the between-response correlation, demonstrates that the joint modeling resulted in improved variable selection. This difference was strongest at low signal-to-noise ratios ($\delta = 0.25, 0.5$). For example, in setting ($\delta = 0.25, \rho = 0.9$) CoMPAdRe had a significantly higher median true positive rate compared to the median rates of its competitors ($\hbox{CoMPAdRe} = 0.770, \hbox{PAdRe}=0.111, \hbox{GAMSEL} = 0.111, \hbox{mSSL}=0.270, \hbox{Lasso} = 0.111$).  
ComPAdRe had the highest sensitivity for all settings except for the highest signal-to-noise ($\delta = 2$), for which GAMSEL had slightly higher sensitivity, but both GAMSEL and COMPAdRe near 1. GAMSEL, however, consistently reported a false positive rate approximately $\sim \hbox{10-fold}$ higher than CoMPAdRe at higher signal to noise ratios ($\delta = [0.75, 1, 2]$) across all levels of residual dependence $\rho$. This pattern persisted across all settings considered, as seen in supplemental section \textbf{S.1}. For example, in settings where ($p = 10, \delta = 2$), GAMSEL showed an average median false positive rate of $\hbox{FPR} = 0.122$ while  CoMPAdRe showed an average median false positive rate of $\hbox{FPR} = 0.011$ across all levels of residual dependence $\rho$.

At lower levels of residual dependence $(\boldsymbol{\rho = 0.2, 0.5})$, CoMPAdRe still generally exhibited superior sensitivity than competitors. The contrast between true positive rates at lower levels of signal to noise ($\delta = [0.25, 0.5]$) was less stark than those seen at higher levels of $\rho$ across all methods. GAMSEL again 
had slightly higher sensitivity than CoMPAdRe at ($\delta =2$) and 
($\delta = 1$), but again accompanied by higher FPR. From additional simulations conducted to assess function-specific selection and estimation performance (supplemental section \textbf{S.1.2}), we see that linear selection approaches (Lasso and mSSL) consistently failed to select functions $f_{2}$ and $f_{4}$, even at high signal-to-noise ratios, while CoMPAdRe while allowing the potential of nonlinear associations did not appear to lose sensitivity for selecting covariates with linear associations.


In terms of estimation accuracy, CoMPAdRe outperformed method \textbf{(1)} when estimating overall signal $\Hat{f}$ across both $\delta$ and $\rho$. In each setting of signal-to-noise $\delta$ the improved performance of CoMPAdRe relative to method $(1)$ increased as level of residual dependence $\rho$ increased; this is evidenced by the downward linear trend in Figure 2 at each level of $\delta$. This trend persisted across all settings considered (see \textbf{S.1}). The only setting where CoMPAdRe didn't show notable gains in estimation accuracy occurred at low signal-to-noise and levels of residual dependence: $(\delta = 0.25, \rho = 0.2), (\delta = 0.25, \rho = 0.5)$. In these settings the ratio of estimation accuracy $(\footnotesize{\frac{\hbox{CoMPAdRe}}{\hbox{PAdRe}}})$ remained centered around 1. 
These results demonstrate the strong benefit in estimation accuracy from the joint modeling when strong inter-response correlation structure is precent, but without substantial tradeoff when the responses have low levels of correlation.
CoMPAdRe displayed similar performance relative to other approaches considered, as seen in supplemental section \textbf{S.1}, with improvements in estimation accuracy increasing with higher levels of inter-response correlation.
From additional function-specific simulations (supplemental section \textbf{S.1.2}), CoMPAdRe demonstrated superior estimation accuracy to linear selection approaches, with results most evident for estimation of $f_{1}$ and $f_{4}$. Given simulation results that show the benefits of our approach, we now apply CoMPAdRe to protein-mRNA expression data obtained from The Cancer Protein Atlas project (TCPA).

\begin{figure}[h]
\centerline{\includegraphics[height=9cm,width=16cm]{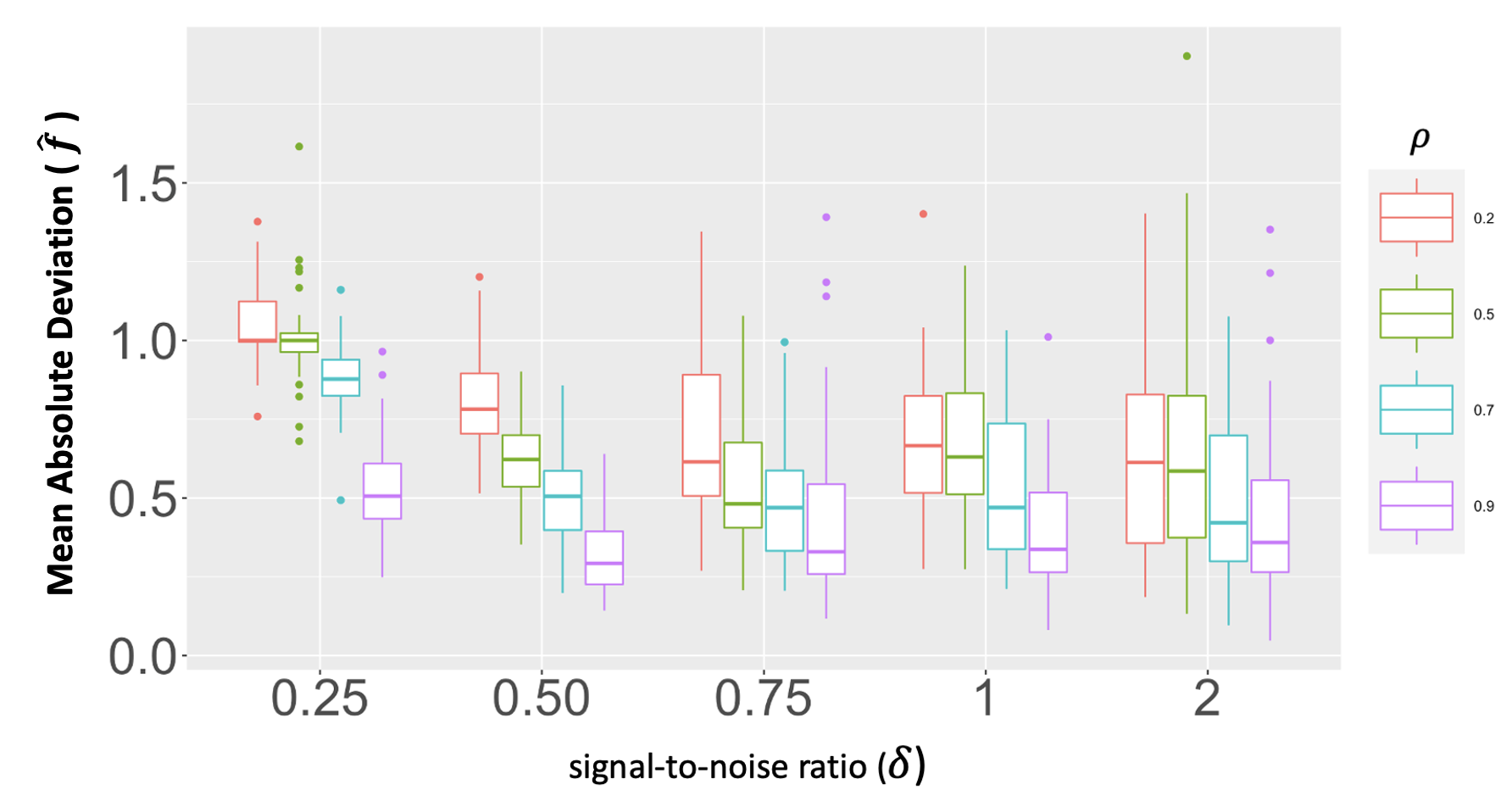}}
\caption{\footnotesize Boxplots of estimation accuracy results for non-linear signal $\Hat{f}$. Midpoint lines represent the median ratio of mean absolute deviation (MAD):  ($\scriptsize{\frac{\hbox{CoMPAdRe}}{\hbox{PAdRe}}}$) across 50 simulated datasets per setting. Settings considered vary based on signal-to-noise ratio $\delta$ and level of residual dependence $\rho$.}\label{fig:01}
\end{figure}

\begin{table}[H]
\vspace{-0.7cm}
\scriptsize
\centering
\setstretch{1.5}
\begin{adjustbox}{angle=90}
\begin{tabular}{|ccccccccccccc|}
\hline
 $\delta$ &  & CoMPAdRe & PAdRe & GAMSEL & mSSL & Lasso &  & CoMPAdRe & PAdRe & GAMSEL & mSSL & Lasso \\ \hline
 \multicolumn{1}{|c}{} & \multicolumn{1}{|c}{$\rho$ = 0.9} &  &  &  &  & \multicolumn{1}{c|}{} & $\rho$ = 0.9 &  &  &  &  &  \\
 0.25 & \multicolumn{1}{|c}{} & 77.0 (21.7) & 11.1 (20.0) & 11.1 (24.3) & 27.0 (21.3) & \multicolumn{1}{c|}{11.1 (20.0)} &  & 0.2 (0.5) & $<$ 0.1 (0) & $<$ 0.1 (0) & $<$ 0.1 (0) & $<$ 0.1 (0) \\
 0.5 & \multicolumn{1}{|c}{} & 87.5 (10.2) & 40.8 (37.1) & 55.8 (34.5) & 71.4 (26.7) & \multicolumn{1}{c|}{40.8 (34.7)} &  & 0.3 (0.6) & $<$ 0.1 (0.2) & 0.7 (0.9) & $<$ 0.1 (0) & $<$ 0.1 (0.2) \\
0.75 & \multicolumn{1}{|c}{TPR} & 93.3 (12.2) & 69.6 (30.7) & 83.3 (19.3) & 73.0 (22.9) & \multicolumn{1}{c|}{64.0 (31.6)} & FPR & 0.4 (0.6) & 0.1 (0.2) & 1.5 (1.9) & $<$ 0.1 (0) & 0.1 (0.2) \\
 1 & \multicolumn{1}{|c}{} & 92.9 (13.3) & 76.9 (16.5) & 91.7 (9.10) & 75.0 (20.7) & \multicolumn{1}{c|}{66.7 (13.5)} &  & 0.2 (0.4) & 0.1 (0.2) & 1.5 (1.1) & $<$ 0.1 (0) & 0.1 (0.2) \\
 2 & \multicolumn{1}{|c}{} & 97.1 (8.20) & 85.1 (12.9) & $>$ 99.8 (0) & 77.8 (15.9) & \multicolumn{1}{c|}{72.7 (19.0)} &  & 0.4 (1.0) & 0.1 (0.2) & 2.5 (1.9) & $<$ 0.1 (0) & 0.1 (0.2) \\ \hline
\multicolumn{1}{|c|}{} & $\rho$ = 0.7 &  &  &  &  & \multicolumn{1}{c|}{} & $\rho$ = 0.7 &  &  &  &  &  \\
 0.25 & \multicolumn{1}{|c}{} & 33.3 (20.8) & 11.1 (24.5) & 10.6 (25.0) & 11.1 (12.3) & \multicolumn{1}{c|}{11.1 (24.5)} &  & 0.1 (0.3) & $<$ 0.1 (0) & $<$ 0.1 (0.1) & $<$ 0.1 (0) & $<$ 0.1 (0) \\
 0.5 & \multicolumn{1}{|c}{} & 83.3 (20.0) & 44.2 (34.6) & 54.5 (23.8) & 55.8 (28.1) & \multicolumn{1}{c|}{43.3 (33.3)} &  & 0.3 (0.5) & $<$ 0.1 (0.2) & 0.6 (1.0) & $<$ 0.1 (0) & $<$ 0.1 (0.2) \\
 \multicolumn{1}{|c|}{0.75} & TPR & 84.6 (13.7) & 60.0 (27.1) & 83.3 (17.5) & 63.6 (22.5) & \multicolumn{1}{c|}{56.3 (22.9)} & FPR & 0.2 (0.3) & 0.1 (0.2) & 1.5 (1.7) & $<$ 0.1 (0) & 0.1 (0.2) \\
 1 & \multicolumn{1}{|c}{} & 90.5 (15.1) & 72.4 (16.3) & 90.0 (18.2) & 66.7 (19.4) & \multicolumn{1}{c|}{63.1 (22.0)} &  & 0.2 (0.4) & 0.1 (0.2) & 1.3 (1.4) & $<$ 0.1 (0) & 0.1 (0.2) \\
 2 & \multicolumn{1}{|c}{} & 94.1 (10.8) & 84.0 (15.7) & $>$ 99.8 (0) & 73.0 (26.2) & \multicolumn{1}{c|}{72.1 (24.8)} &  & 0.3 (0.6) & $<$ 0.1 (0.3) & 2.8 (1.8) & $<$ 0.1 (0) & $<$ 0.1 (0.3) \\ \hline
 \multicolumn{1}{|c|}{} & $\rho$ = 0.5 &  &  &  &  & \multicolumn{1}{c|}{} & $\rho$ = 0.5 &  &  &  &  &  \\
 0.25 & \multicolumn{1}{|c}{} & 15.4 (21.5) & 10.6 (19.7) & 12.5 (17.1) & 11.1 (12.7) & \multicolumn{1}{c|}{10.6 (19.7)} &  & $<$ 0.1 (0.3) & $<$ 0.1 (0) & $<$ 0.1 (0) & $<$ 0.1 (0) & $<$ 0.1 (0) \\
 0.5 & \multicolumn{1}{|c}{} & 72.7 (16.7) & 44.9 (26.3) & 56.9 (23.3) & 33.3 (19.4) & \multicolumn{1}{c|}{44.4 (27.8)} &  & 0.3 (0.1) & 0.1 (0.2) & 0.6 (1.0) & $<$ 0.1 (0) & 0.1 (0.2) \\ 0.75
 & \multicolumn{1}{|c}{TPR} & 84.0 (19.1) & 64.2 (29.8) & 81.8 (22.2) & 65.2 (25.1) & \multicolumn{1}{c|}{61.5 (29.1)} & FPR & 0.2 (0.2) & 0.1 (0.3) & 1.6 (1.6) & $<$ 0.1 (0) & 0.1 (0.3) \\
 1 & \multicolumn{1}{|c}{} & 88.9 (15.4) & 78.2 (23.9) & 89.4 (20.0) & 66.7 (28.4) & \multicolumn{1}{c|}{69.2 (23.5)} &  & 0.2 (0.3) & $<$ 0.1 (0.2) & 1.5 (1.5) & $<$ 0.1 (0) & $<$ 0.1 (0.2) \\
 2 & \multicolumn{1}{|c}{} & 92.9 (14.0) & 84.0 (17.2) & $>$ 99.8 (0) & 72.7 (15.7) & \multicolumn{1}{c|}{75.0 (19.0)} &  & 0.2 (0.4) & 0.1 (0.1) & 3.0 (2.5) & $<$ 0.1 (0) & 0.1 (0.1) \\ \hline
 & \multicolumn{1}{|c}{$\rho$ = 0.2}   &  &  &  &  & \multicolumn{1}{c|}{} & $\rho$ = 0.2 &  &  &  &  &  \\
 0.25 & \multicolumn{1}{|c}{} & 16.2 (24.3) & 12.5 (19.5) & 19.1 (19.8) & 11.1 (12.2) & \multicolumn{1}{c|}{12.5 (19.5)} &  & $<$ 0.1 (0.1) & $<$ 0.1 (0) & $<$ 0.1 (0.1) & $<$ 0.1 (0) & $<$ 0.1 (0) \\
 0.5 & \multicolumn{1}{|c}{} & 57.3 (20.2) & 36.9 (27.5) & 64.3 (23.2) & 23.1 (20.8) & \multicolumn{1}{c|}{36.9 (25.0)} &  & 0.1 (0.3) & $<$ 0.1 (0.2) & 0.8 (1.3) & $<$ 0.1 (0) & $<$ 0.1 (0.2) \\ \multicolumn{1}{|c|}{0.75}
 & TPR & 84.6 (17.3) & 70.7 (28.9) & 81.5 (21.5) & 57.1 (18.6) & \multicolumn{1}{c|}{65.5 (27.5)} & FPR & 0.2 (0.4) & $<$ 0.1 (0.2) & 1.3 (1.4) & $<$ 0.1 (0) & $<$ 0.1 (0.2) \\
 1 & \multicolumn{1}{|c}{} & 83.3 (17.9) & 72.1 (23.9) & 90.0 (20.0) & 64.5 (26.1) & \multicolumn{1}{c|}{63.1 (28.0)} &  & 0.2 (0.4) & 0.1 (0.2) & 1.5 (1.1) & $<$ 0.1 (0) & 0.1 (0.2) \\
 2 & \multicolumn{1}{|c}{} & 93.3 (11.1) & 84.6 (20.2) & $>$ 99.8 (0) & 72.1 (21.8) & \multicolumn{1}{c|}{72.7 (21.3)} &  & 0.2 (0.6) & 0.1 (0.2) & 2.4 (1.6) & $<$ 0.1 (0) & 0.1 (0.2) \\ \hline
\end{tabular}
\end{adjustbox}
\caption{\tiny Summary of simulation results for settings where number of covariates $p = 100$. Results are divided into true positive rate (TPR) and false positive rate (FPR), expressed as a percentage, for levels of residual dependence $\rho = (0.2, 0.5, 0.7, 0.9)$ and signal-to-noise ratio $\delta = (0.25, 0.5, 0.75, 1, 2)$. Signal is divided into null vs. non-null signal and results are presented as the median with the interquartile range in parenthesis. 50 datasets were simulated for each setting and sample size was fixed to $n = 250$ and the number of responses to $q = 10$.}
\end{table}

\section{Analyses of Proteogenomics data in Breast Cancer}
We applied CoMPAdRe to a proteogenomics dataset containing both protein and mRNA expression levels for 8 known breast cancer pathways obtained from The Cancer Protein Atlas project (TCPA). The central dogma of molecular biology states that genetic information is primarily passed from an individual's DNA towards the production of mRNA (a process called transcription) before mRNA takes that information to a cell's ribosomes to construct proteins (a process called translation) which carry out the body's biological functions \citep{crick1970central}. Furthermore, proteins are known to carry out biological functions in coordinated networks \citep{de2010protein, garcia2012networks}. Here we consider a subset of TCPA data consisting of mRNA and protein data for $n=844$ subjects from $8$ cancer-relevant biological pathways.  For each pathway, we have data from $3-11$ proteins and 3 $-$ 11 mRNA transcripts.  Our objective is three-fold: (1) to find which mRNA are predictive of protein expression in a particular pathway, (2) assess the shape of the relationship, whether linear or nonlinear, and (3) estimate the protein-protein networks accounting for mRNA expression.  We will accomplish this by applying CoMPAdRe to each pathway, treating the $Q$ proteins as responses and $p$ mRNA transcripts as covariates.  The joint approach will give us estimates of the $Q \times Q$ protein-protein network and, as demonstrated by our simulations, we expect this joint modeling will result in improved detection of mRNA transcripts predictive of protein abundances than if the proteins were modeled independently.

More information about each pathway can be found in \citep{akbani2014pan,ha2018personalized}. As an initial step, covariates (mRNA expression levels) were centered and scaled by their mean and standard deviation.  Just as in our simulation studies, we then formed B-spline bases by taking knots at the deciles for each covariate considered within a pathway. 

\paragraph{Biological Interpretations:} Table 2 summarizes mRNA selection results for all pathways analyzed. A comprehensive table of mRNA selection results  are contained in supplemental section \textbf{S.2} along with visualizations of the shapes for selected nonlinear mRNA--protein associations. Among non-linear associations, we note that CDH1--$\beta$-catenin and CDH1--E-cadherin were selected for the Core Reactive Pathway, CDH1--E-cadherin for the EMT pathway, INPP4B--INPP4B for the PI3K-AKT pathway, and ERBB2--HER2PY1248 for the RTK pathway. We consistently found the same pattern for all non-linear mRNA--protein associations; protein expression increases with mRNA expression until seemingly hitting a plateau. Protein expression has been found to both be positively correlated with mRNA expression and to often plateau at high expression levels because of a suspected saturation of ribosomes, which would limit translation \citep{liu2016dependency, van2021protein}. Therefore, identified non-linear mRNA--protein functional associations may provide insight into `saturation points' beyond which increased mRNA expression ceases to result in an increase of protein abundance.

\begin{table}[h]
\centering
\setstretch{1.9}
\footnotesize
\begin{tabular}{|c|c|c|}
\hline
Pathway         & Linear mRNA                                  & Non-linear mRNA        \\ \hline
Breast Reactive & GAPDH-GAPDH                                                      & -      \\
Core Reactive   & -                                                             & CDH1--$\beta$-catenin, ... \\
DNA damage      & RAD50-RAD50, MRE11A-RAD50, ...              & -                             \\
EMT             & CDH1--$\beta$-catenin                                                 & CDH1--E-cadherin                   \\
PI3K - AKT & PTEN-PTEN, CDK1B-AKTPT308, ... & INPP4B-INPP4B \\
RAS-MAPK        & YBX1-JNKPT183Y185,  YBX1-YB1PS102                                & -                             \\
RTK             & ERBB2-EGFRPY1068, EGFR-EGFRPY1068                                & ERBB2-HER2PY1248                 \\
TSC-mTOR        & EIF4EBP1-X4EBP1PS65, ... & -                             \\ \hline
\end{tabular}
\caption{\footnotesize A summary of mRNA selection results for each pathway, divided into linear associations and non-linear associations. Names on the left side of the dash indicate mRNA biomarkers, while those on the right side of the dash indicate the corresponding matching protein. Lines ended by '...' indicate the presence of one or more associations than those listed.}
\end{table}

\begin{figure}[H]
\centerline{\includegraphics[height=13cm,width=\columnwidth]{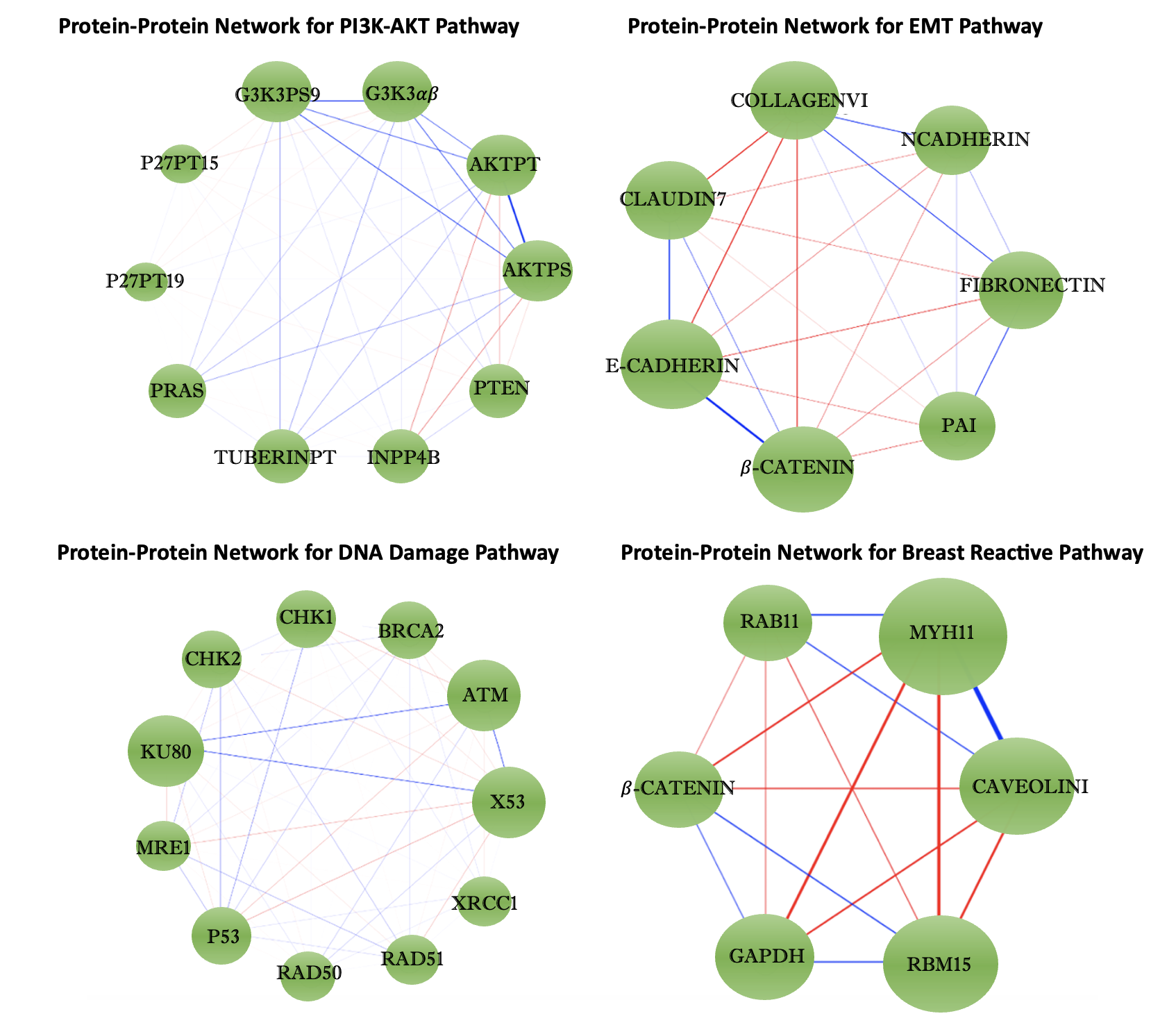}}
\centering
\caption{\setstretch{1.1}\footnotesize Protein-Protein covariance networks for PI3K-AKT, EMT, DNA Damage, and Breast Reactive pathways. Blue edges indicate negative associations while red edges indicate positive associations. Edge thickness indicates the magnitude of the dependence between two corresponding proteins and node size is scaled relative to the strength and number of connections for a protein.}

\end{figure}

\vspace{-0.5cm}
\paragraph{Pathway specific protein-protein networks:} Figure 2 displays protein-protein networks for the pathways detailed in mRNA selection. A visualization of protein-protein networks for all other pathways are detailed in supplemental section \textbf{S.2}. The most highly interconnected proteins, which we call hub nodes, include MYH11, E-cadherin, $\beta$-catenin, and EGFR. In particular, MYH11 showed strong associations with other proteins in the Breast Reactive pathway. MYH11 are smooth muscle myosin heavy chain proteins, which play an essential role in   cell movement and the transport of materials within and between cells \citep{brownstein2018genetic}. While its exact function in breast cancer is not fully understood, MYH11 has been found to be downregulated in breast cancer tissues and to have been critical to empirically constructed indicators for survival prognosis in breast cancer \citep{zhu2020construction}. Our results indicate that the role of MYH11 in breast cancer may be better understood through the biomarkers it was found to be highly associated with (CAVEOLINI, RBM15, GAPDH). We also note that previous literature has shown expression levels of E-cadherin and $\beta$-catenin to be strongly correlated with each other; reduced levels in both proteins are associated with poor survival prognosis in triple negative breast cancer \citep{shen2016prognostic}.

\section{Discussion}
In this article, we introduced CoMPAdRe, a framework for simultaneous variable selection and estimation of sparse residual precision matrices in multivariate additive models.  The approach simultaneously obtains a sparse estimate of the inter-response precision matrix and utilizes this association to borrow strength across responses in identifying significant covariates for each response, determining whether each selected covariate effect is linear or nonlinear.  It also obtains an estimate of the residual precision matrix, which itself is a quantity of scientific interest in many settings including the protein-protein network modeling of our motivating example.  Importantly, our approach allows different covariates to be selected for different responses, and the same covariate to have different functional relationships for different responses.  The fitting procedure utilizes a regression approach to estimate and account for the residual correlation that simplifies the joint multivariate modeling into a series of single-response models to improve scalability and computational efficiency.  We also note that the univariate special case of our method, which we call \textit{PAdRe}, is a useful tool for doing linear and nonlinear variable selection in univariate additive model setting, as well.

We empirically demonstrated that CoMPAdRe achieves superior overall variable selection accuracy and statistical efficiency than marginal competitors that model each response independently, with the benefit more demonstrable as the inter-response correlation increases or the signal-to-noise levels decrease.  This demonstrates the benefit of joint multivariate modeling, with the inherent borrowing of strength across responses resulting in not just better estimation accuracy, but also improved variable selection and determination of linearity or nonlinearity of the effects. 
The improved estimation accuracy was expected based on the principles of \textit{seemingly unrelated regression} (SUR, \cite{zellner1963estimators}), given that the sparsity prior on covariates implies different covariates for each response.  However, our results also suggest that the joint modeling cansubstantially improve variable selection accuracy, as well.  Future theoretical investigations would be interesting to evaluate and validate this result.  Further, the improved performance for lower signal-to-noise ratios demonstrates that the efficiency gained by borrowing strength across responses may be especially important to detect more subtle signals or in the presence of higher noise levels. Our simulations also demonstrate that for covariates with nonlinear associations with responses, variable selection methods based on linear regression will tend to miss these variables, especially for certain nonlinear functional shapes.  Thus, the variable selection framework we introduce enables identification of important predictive variables even when having a highly nonlinear association with responses.


While the CoMPAdRe method introduced here assumes unstructured inter-response correlation structure, it would be relatively straightforward to adapt to incorporate structured covariance matrices, for example if the responses are observed on some known temporal or spatial grid, or if some decomposible graph structure is known beforehand.  This extension could bring the enhanced variable selection and estimation accuracy inherent to the joint modeling while accounting for known structure among the responses.  We leave this work for future extensions.  This paper focused on variable selection and estimation, but it would also be of interest to obtain inferential quantities, including hypothesis tests for significant covariates or confidence measures for each selected variable as well as confidence bands for estimated regression functions and precision elements.  We also leave this to future work.
Finally, while as described in the supplement we have semi-automatic methods to estimate the tuning parameters that seem to work well, more rigorous and automatic approaches for selecting smoothing parameters in this setting may improve performance further, and will be investigated in the future. Software related to this article can be found at \url{https://github.com/nmd1994/ComPAdRe} along with examples for implementation.


\section{Acknowledgements}
This work was partially supported by CA-178744 and CA-244845 from the National Cancer Institute, and TR-001878 from the National Center for Advancing Translational Science.

\bibliographystyle{agsm}
\small
\setstretch{1.2}
\bibliography{Bibliography-MM-MC.bib}

\section*{Supplementary Materials}
\subsection*{S.1: Simulation Results}
We begin this supplemental section by displaying estimation accuracy results, where CoMPAdRe's performance is compared to the other methods considered (GAMSEL, Lasso, and the mSSL). We see the same trends observed in the main body of this manuscript, where CoMPAdRe's improved performance relative to other approaches increases as $\rho$ is increased. We note that GAMSEL does not re-estimate post selection as CoMPAdRe, PAdRe, mSSL, and the Lasso (where we re-estimated post-selection as often recommended in practice). This property likely contributes negatively to the method's estimation performance.

\vspace{0.2cm}

\hspace{-0.75cm} \large \textbf{Estimation Accuracy Results for $\boldsymbol{p =100}$:}

\begin{figure}[H]
\centerline{\includegraphics[height=8cm,width=15cm]{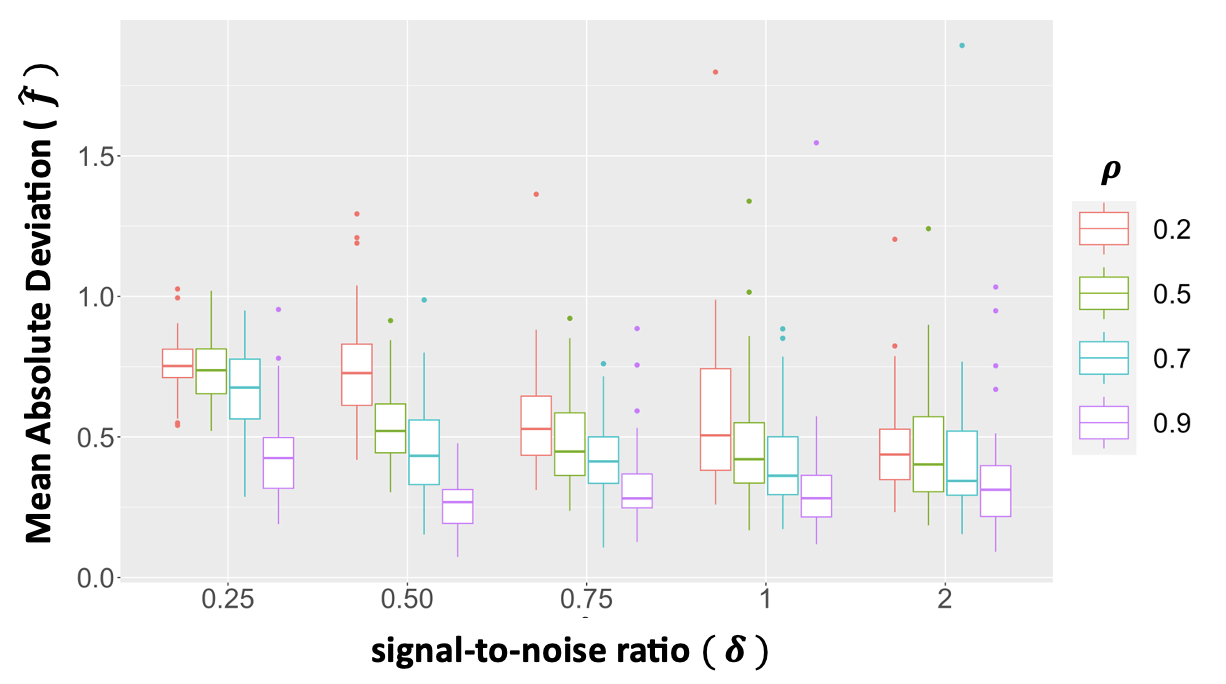}}
\caption{\footnotesize Boxplots of estimation accuracy results for non-linear signal $\Hat{f}$. Midpoint lines represent the median ratio of mean absolute deviation (MAD):  ($\scriptsize{\frac{\hbox{CoMPAdRe}}{\hbox{GAMSEL}}}$) across 50 simulated datasets per setting. Settings considered vary based on signal-to-noise ratio $\delta$ and level of residual dependence $\rho$.}\label{fig:01}
\end{figure}

\begin{figure}[H]
\centerline{\includegraphics[height=8cm,width=15cm]{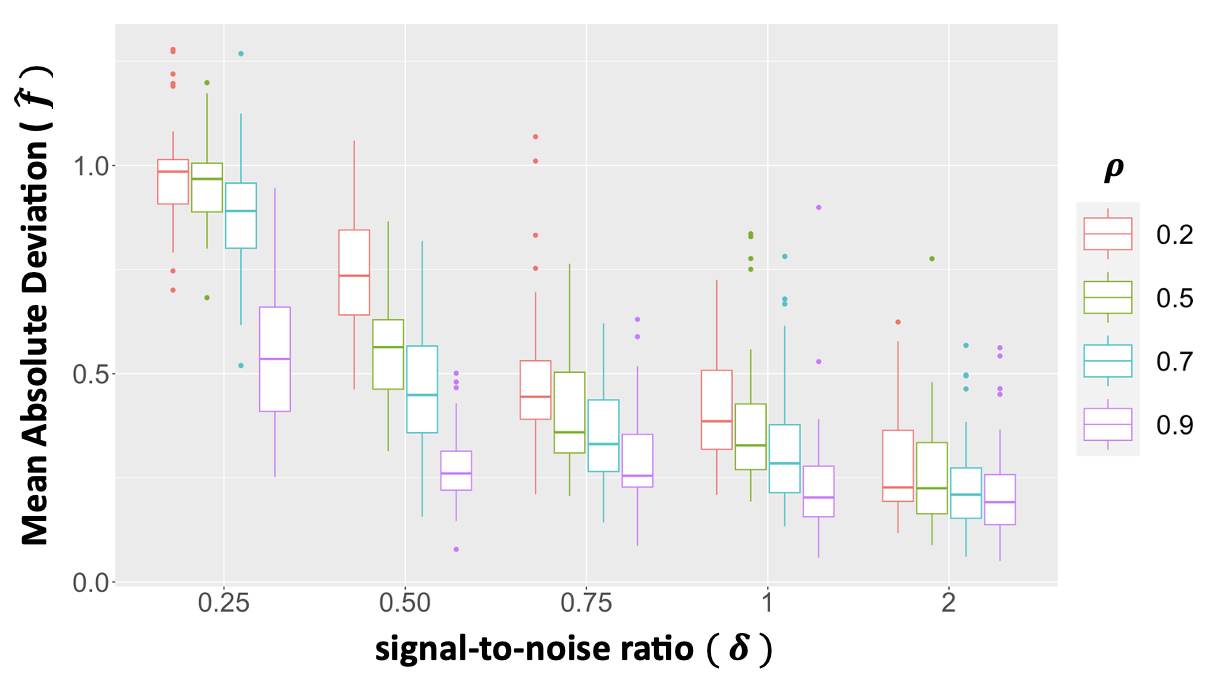}}
\caption{\footnotesize Boxplots of estimation accuracy results for non-linear signal $\Hat{f}$. Midpoint lines represent the median ratio of mean absolute deviation (MAD):  ($\scriptsize{\frac{\hbox{CoMPAdRe}}{\hbox{Lasso}}}$) across 50 simulated datasets per setting. Settings considered vary based on signal-to-noise ratio $\delta$ and level of residual dependence $\rho$.}\label{fig:01}
\end{figure}

\begin{figure}[H]
\centerline{\includegraphics[height=8cm,width=15cm]{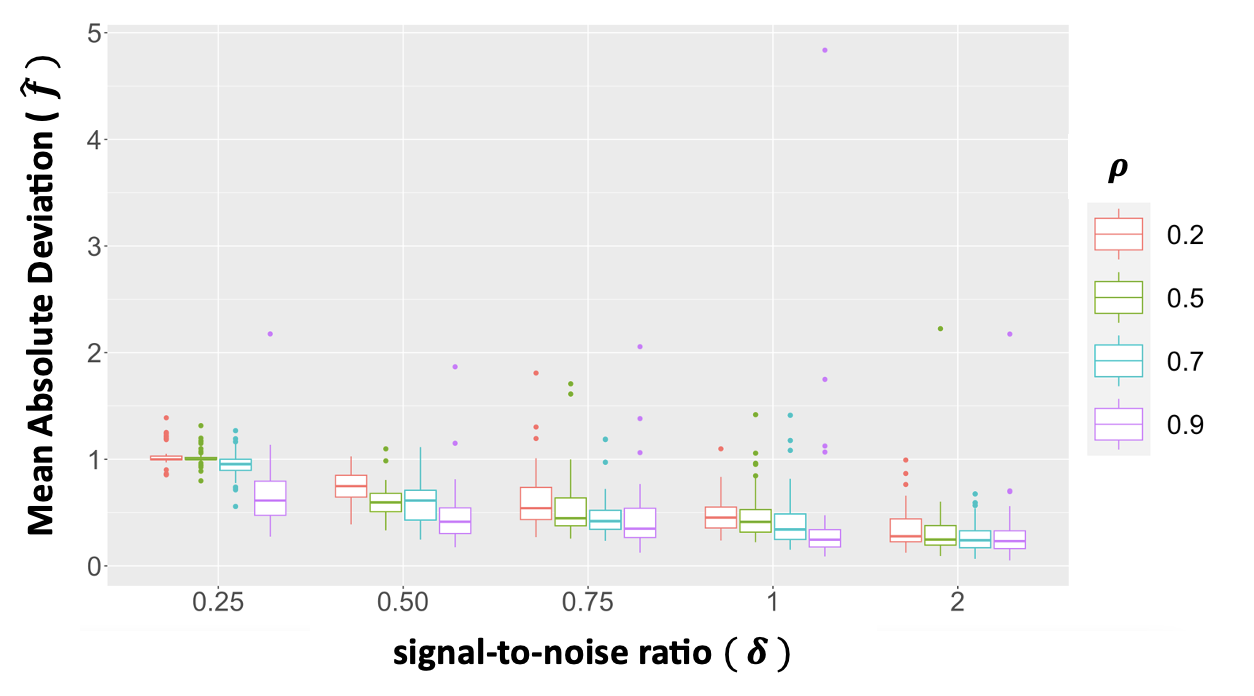}}
\caption{\footnotesize Boxplots of estimation accuracy results for non-linear signal $\Hat{f}$. Midpoint lines represent the median ratio of mean absolute deviation (MAD):  ($\scriptsize{\frac{\hbox{CoMPAdRe}}{\hbox{mSSL}}}$) across 50 simulated datasets per setting. Settings considered vary based on signal-to-noise ratio $\delta$ and level of residual dependence $\rho$.}\label{fig:01}
\end{figure}

\newpage
We next show selection results subdivided by function type (linear, non-linear) for the simulations presented in the main body of this manuscript (i.e. settings where $p = 100$). 
\newline
\newline
\noindent \large \textbf{Linear Selection Results for $\boldsymbol{p =100}$:}

\begin{table}[H]
\centering
\small
\begin{tabular}{|cccccc|}
\hline
 &                           &             &             $\rho = 0.9$                &    &                    \\
 & $\delta$                     & CoMPAdRe      & PAdRe & GAMSEL    &        \\ \hline
    & \multicolumn{1}{c|}{0.25} & $>$ 99.7 (20.0) & $<$ 0.30 (16.1) & $<$ 0.30 (15.6) & \\
    & \multicolumn{1}{c|}{0.5}  & $>$ 99.7 (0) & 53.6 (44.2) & 75.0 (50.0) & \\
TPR & \multicolumn{1}{c|}{0.75} & $>$ 99.7 (0) & $>$ 99.7 (14.3) & $>$ 99.7 (0) & \\
    & \multicolumn{1}{c|}{1}    & $>$ 99.7 (0) & $>$ 99.7 (14.3) & $>$ 99.7 (0) & \\
 & \multicolumn{1}{c|}{2}    & $>$ 99.7 (0) & $>$ 99.7 (0) &  $>$ 99.7 (0) & \\ \hline
 & \multicolumn{1}{c|}{0.25} & 0.10 (0.50) & $<$ 0.10 (0) &  $<$ 0.10 (0) &  \\
 & \multicolumn{1}{c|}{0.5}  & 0.20 (0.70) & $<$ 0.10 (0.20) & 0.60 (0.90) &   \\
FPR & \multicolumn{1}{c|}{0.75} & 0.30 (0.50) & 0.10 (0.20) & 1.20 (1.70) &  \\
 & \multicolumn{1}{c|}{1}    & 0.20 (0.40) & 0.10 (0.20) & 1.00 (1.00) &  \\
 & \multicolumn{1}{c|}{2}    & 0.30 (0.80) & 0.10 (0.20) & 2.20 (1.80) & \\ \hline
\end{tabular}
\end{table}

\begin{table}[H]
\centering
\small
\begin{tabular}{|cccccc|}
\hline
 &                           &             &             $\rho = 0.7$                &    &                    \\
 & $\delta$                     & CoMPAdRe      & PAdRe & GAMSEL    &        \\ \hline
    & \multicolumn{1}{c|}{0.25} & 31.7 (50.0) & $<$ 0.30 (0)  & $<$ 0.30 (0) & \\
    & \multicolumn{1}{c|}{0.5}  & $>$ 99.7 (11.1) & 50.0 (41.7) & 75.0 (27.1) & \\
TPR & \multicolumn{1}{c|}{0.75} & $>$ 99.7 (0) & 87.5 (23.8) & $>$ 99.7 (0) & \\
    & \multicolumn{1}{c|}{1}    &  $>$ 99.7 (0) & $>$ 99.7 (20.0) & $>$ 99.7 (0) & \\
 & \multicolumn{1}{c|}{2}    & $>$ 99.7 (0) & $>$ 99.7 (0) &  $>$ 99.7 (0) & \\ \hline
 & \multicolumn{1}{c|}{0.25} & 0.10 (0.20) & $<$ 0.10 (0) & $<$ 0.10 (0.10)  &  \\
 & \multicolumn{1}{c|}{0.5}  & 0.20 (0.30) & $<$ 0.10 (0.20) & 0.70 (0.90) &   \\
FPR & \multicolumn{1}{c|}{0.75} & 0.20 (0.30) & 0.10 (0.20) & 1.20 (1.50) &  \\
 & \multicolumn{1}{c|}{1}    & 0.20 (0.40) & 0.10 (0.20) & 0.90 (1.30) &  \\
 & \multicolumn{1}{c|}{2}    & 0.30 (0.50) & $<$ 0.10 (0.30) & 2.30 (1.70) & \\ \hline
\end{tabular}
\end{table}

\begin{table}[H]
\centering
\small
\begin{tabular}{|cccccc|}
\hline
 &                           &             &             $\rho = 0.5$                &    &                    \\
 & $\delta$                     & CoMPAdRe      & PAdRe & GAMSEL    &        \\ \hline
    & \multicolumn{1}{c|}{0.25} & 5.00 (25.0) & $<$ 0.30  (0)  & $<$ 0.30 (15.1) & \\
    & \multicolumn{1}{c|}{0.5}  & 89.4 (24.3) & 60.0 (31.3) & 80.0 (33.3) & \\
TPR & \multicolumn{1}{c|}{0.75} & $>$ 99.7 (0) & 94.4 (23.8) & $>$ 99.7 (0) & \\
    & \multicolumn{1}{c|}{1}    & $>$ 99.7 (0) & $>$ 99.7 (11.9) & $>$ 99.7 (0) & \\
 & \multicolumn{1}{c|}{2}    & $>$ 99.7 (0) & $>$ 99.7 (0) &  $>$ 99.7 (0) & \\ \hline
 & \multicolumn{1}{c|}{0.25} & $<$ 0.10 (0.10) & $<$ 0.10 (0) & $<$ 0.10 (0)  &  \\
 & \multicolumn{1}{c|}{0.5}  & 0.20 (0.40) & 0.10 (0.20) & 0.50 (0.80) &   \\
FPR & \multicolumn{1}{c|}{0.75} & 0.10 (0.20) & 0.10 (0.30) & 1.30 (1.40) &  \\
 & \multicolumn{1}{c|}{1}    & 0.10 (0.30) & $<$ 0.10 (0.20) & 1.30 (1.50) &  \\
 & \multicolumn{1}{c|}{2}    & 0.20 (0.30) & 0.10 (0.10) & 2.60 (2.30) & \\ \hline
\end{tabular}
\end{table}

\begin{table}[H]
\centering
\small
\begin{tabular}{|cccccc|}
\hline
 &                           &             &             $\rho = 0.2$                &    &                    \\
 & $\delta$                     & CoMPAdRe      & PAdRe & GAMSEL    &        \\ \hline
    & \multicolumn{1}{c|}{0.25} & $<$ 0.30  (14.3) & $<$ 0.30  (0) & $<$ 0.30  (14.3) & \\
    & \multicolumn{1}{c|}{0.5}  & 52.3 (41.8) & 46.4 (43.4) & 78.9 (50.0) & \\
TPR & \multicolumn{1}{c|}{0.75} & $>$ 99.7 (13.5) & 88.9 (32.1) & $>$ 99.7 (0) & \\
    & \multicolumn{1}{c|}{1}    & $>$ 99.7 (0) & $>$ 99.7 (20.0) & $>$ 99.7 (0) & \\
 & \multicolumn{1}{c|}{2}    & $>$ 99.7 (0) & $>$ 99.7 (0) &  $>$ 99.7 (0) & \\ \hline
 & \multicolumn{1}{c|}{0.25} & $<$ 0.10 (0) & $<$ 0.10 (0) & $<$ 0.10 (0.10)  &  \\
 & \multicolumn{1}{c|}{0.5}  & $<$ 0.10 (0.20) & $<$ 0.10 (0.20) & 0.70 (1.10) &   \\
FPR & \multicolumn{1}{c|}{0.75} & 0.10 (0.40) & $<$ 0.10 (0.20) & 1.10 (1.3) &  \\
 & \multicolumn{1}{c|}{1}    & 0.10 (0.40) & 0.10 (0.20) & 1.20 (1.30) &  \\
 & \multicolumn{1}{c|}{2}    & 0.10 (0.40) & 0.10 (0.20) & 2.10 (1.50) & \\ \hline
\end{tabular}
\caption{\textbf{Linear Selection Results: }Summary of selection results for linear functions for methods that select both linear and non-linear functions in settings where number of covariates $p = 100$. Results are divided into true positive rate (TPR) and false positive rate (FPR), expressed as a percentage, for levels of residual dependence $\rho = (0.2, 0.5, 0.7, 0.9)$ and signal-to-noise ratio $\delta = (0.25, 0.5, 0.75, 1, 2)$. Results are presented as the median with the interquartile range in parenthesis. 50 datasets were simulated for each setting and sample size was fixed to $n = 250$ and the number of responses to $q = 10$.}
\end{table}

\newpage

\noindent \large \textbf{Non-Linear Selection Results for $\boldsymbol{p =100}$:}

\begin{table}[H]
\centering
\small
\begin{tabular}{|cccccc|}
\hline
 &                           &             &             $\rho = 0.9$                &    &                    \\
 & $\delta$                     & CoMPAdRe      & PAdRe & GAMSEL    &        \\ \hline
    & \multicolumn{1}{c|}{0.25} & 37.5 (30.0) & $<$ 0.30 (0) & $<$ 0.30 (0) & \\
    & \multicolumn{1}{c|}{0.5}  & 55.6 (26.0) & $<$ 0.30 (0) & 22.2 (23.4) & \\
TPR & \multicolumn{1}{c|}{0.75} & 55.6 (31.0) & 10.0 (20.0) & 40.0 (30.0) & \\
    & \multicolumn{1}{c|}{1}    & 66.7 (30.0) & 33.3 (27.5) & 66.7 (32.1) & \\
 & \multicolumn{1}{c|}{2}    & 71.4 (21.9) & 33.3 (24.1) &  75.0 (37.2)  & \\ \hline
 & \multicolumn{1}{c|}{0.25} & $<$ 0.10 (0.10) & $<$ 0.10 (0) & 0.10 (0.20)  &  \\
 & \multicolumn{1}{c|}{0.5}  & $<$ 0.10 (0.10) & $<$ 0.10 (0) & 0.10 (0.20) &   \\
FPR & \multicolumn{1}{c|}{0.75} & $<$ 0.10 (0.10) & $<$ 0.10 (0) & $<$ 0.10 (0.10) &  \\
 & \multicolumn{1}{c|}{1}    & $<$ 0.10 (0) & 0.20 (0.30) & $<$ 0.10 (0) &  \\
 & \multicolumn{1}{c|}{2}    & $<$ 0.10 (0.10) & $<$ 0.10 (0) & 0.30 (0.30) & \\ \hline
\end{tabular}
\end{table}

\begin{table}[H]
\centering
\small
\begin{tabular}{|cccccc|}
\hline
 &                           &             &             $\rho = 0.7$                &    &                    \\
 & $\delta$                     & CoMPAdRe      & PAdRe & GAMSEL    &        \\ \hline
    & \multicolumn{1}{c|}{0.25} & $<$ 0.30 (14.3) & $<$ 0.30 (0) & $<$ 0.30 (0) & \\
    & \multicolumn{1}{c|}{0.5}  & 40.0 (36.9) & $<$ 0.30 (0) & 33.3 (28.9) & \\
TPR & \multicolumn{1}{c|}{0.75} & 50.0 (33.3) &  12.5 (20.0) & 50.0 (22.2) & \\
    & \multicolumn{1}{c|}{1}    & 50.0 (21.0) & 25.0 (25.1) & 50.0 (22.5) & \\
 & \multicolumn{1}{c|}{2}    & 68.3 (23.3) & 33.3 (23.6) &  77.8 (22.2)  & \\ \hline
 & \multicolumn{1}{c|}{0.25} & $<$ 0.10 (0) & $<$ 0.10 (0) & $<$ 0.10 (0)  &  \\
 & \multicolumn{1}{c|}{0.5}  & $<$ 0.10 (0.20) & $<$ 0.10 (0) & 0.10 (0.20) &   \\
FPR & \multicolumn{1}{c|}{0.75} & $<$ 0.10 (0.10) & $<$ 0.10 (0) & 0.20 (0.20) &  \\
 & \multicolumn{1}{c|}{1}    & $<$ 0.10 (0) & $<$ 0.10 (0) & 0.20 (0.20) &  \\
 & \multicolumn{1}{c|}{2}    & $<$ 0.10 (0) & $<$ 0.10 (0) & 0.40 (0.50) & \\ \hline
\end{tabular}
\end{table}

\begin{table}[H]
\centering
\small
\begin{tabular}{|cccccc|}
\hline
 &                           &             &             $\rho = 0.5$                &    &                    \\
 & $\delta$                     & CoMPAdRe      & PAdRe & GAMSEL    &        \\ \hline
    & \multicolumn{1}{c|}{0.25} & $<$ 0.30 (0) & $<$ 0.30 (0) & $<$ 0.30 (0) & \\
    & \multicolumn{1}{c|}{0.5}  & 42.9 (28.6) & $<$ 0.30 (0) & 22.5 (37.2) & \\
TPR & \multicolumn{1}{c|}{0.75} & 52.8 (32.3) & 15.5 (25.0) & 50.0 (39.0) & \\
    & \multicolumn{1}{c|}{1}    & 50.0 (26.7) & 33.3 (30.0) & 52.3 (29.2) & \\
 & \multicolumn{1}{c|}{2}    & 66.7 (21.6) & 40.0 (21.4) &  75.0 (30.8)  & \\ \hline
 & \multicolumn{1}{c|}{0.25} & $<$ 0.10 (0) & $<$ 0.10 (0) & $<$ 0.10 (0)  &  \\
 & \multicolumn{1}{c|}{0.5}  & $<$ 0.10 (0.10) & $<$ 0.10 (0) & 0.10 (0.20) &   \\
FPR & \multicolumn{1}{c|}{0.75} & $<$ 0.10 (0.10) & $<$ 0.10 (0) & 0.20 (0.20) &  \\
 & \multicolumn{1}{c|}{1}    & $<$ 0.10 (0) & $<$ 0.10 (0) & 0.20 (0.20) &  \\
 & \multicolumn{1}{c|}{2}    & $<$ 0.10 (0) & $<$ 0.10 (0) &  0.40 (0.30) & \\ \hline
\end{tabular}
\end{table}

\begin{table}[H]
\centering
\small
\begin{tabular}{|cccccc|}
\hline
 &                           &             &             $\rho = 0.2$                &    &                    \\
 & $\delta$                     & CoMPAdRe      & PAdRe & GAMSEL    &        \\ \hline
    & \multicolumn{1}{c|}{0.25} & $<$ 0.30 (0) & $<$ 0.30 (0) & $<$ 0.30 (0) & \\
    & \multicolumn{1}{c|}{0.5}  & 29.3 (33.3) & $<$ 0.30 (0) & 29.3 (33.9) & \\
TPR & \multicolumn{1}{c|}{0.75} & 44.4 (40.4) & 10.6 (25.0) & 42.9 (40.0) & \\
    & \multicolumn{1}{c|}{1}    & 50.0 (29.2) & 25.0 (27.1) & 60.0 (31.4) & \\
 & \multicolumn{1}{c|}{2}    & 66.7 (31.9) & 35.4 (25.0) &  80.0 (33.3)  & \\ \hline
 & \multicolumn{1}{c|}{0.25} & $<$ 0.10 (0) & $<$ 0.10 (0) & $<$ 0.10 (0) &  \\
 & \multicolumn{1}{c|}{0.5}  & $<$ 0.10 (0.10) & $<$ 0.10 (0) & 0.10 (0.20) &   \\
FPR & \multicolumn{1}{c|}{0.75} & $<$ 0.10 (0.10) & $<$ 0.10 (0) & 0.20 (0.20) &  \\
 & \multicolumn{1}{c|}{1}    & $<$ 0.10 (0) & $<$ 0.10 (0) & 0.30 (0.10) &  \\
 & \multicolumn{1}{c|}{2}    & $<$ 0.10 (0.10) & $<$ 0.10 (0) &  0.40 (0.40) & \\ \hline
\end{tabular}
\caption{\textbf{Non-Linear Selection Results: }Summary of selection results for Non-Linear functions for methods that select both Linear and Non-Linear functions in settings where number of covariates $p = 100$. Results are divided into true positive rate (TPR) and false positive rate (FPR), expressed as a percentage, for levels of residual dependence $\rho = (0.2, 0.5, 0.7, 0.9)$ and signal-to-noise ratio $\delta = (0.25, 0.5, 0.75, 1, 2)$. Results are presented as the median with the interquartile range in parenthesis. 50 datasets were simulated for each setting and sample size was fixed to $n = 250$ and the number of responses to $q = 10$.}
\end{table}

\newpage
\noindent We next present results for simulated data where the number of covariates is set to $p = 10$. We first present the same selection and estimation results shown in the main body of this manuscript before subdividing into results for linear and non-linear functions. 
\newline 
\newline
\noindent \large \textbf{Results for $\boldsymbol{p =10}$:}

\begin{table}[H]
\centering
\small
\begin{tabular}{|ccccccc|}
\hline
 &                           &               &                             & \hspace{-4cm}$\rho = 0.9$     &               &               \\
 & $\delta$                     & CoMPAdRe      & PAdRe & GAMSEL        & mSSL          & Lasso         \\ \hline
    & \multicolumn{1}{c|}{0.25} & 76.0 (20.5) & 19.1 (15.0) & 19.4 (26.8) & 36.9 (28.7) & 13.4 (17.9) \\
    & \multicolumn{1}{c|}{0.5}  & 92.8 (13.8) & 66.7 (24.4) & 78.6 (20.3) & 66.7 (25.3) & 50.0 (29.7) \\
TPR & \multicolumn{1}{c|}{0.75} & $>$ 99.8 (8.30)   & 85.2 (15.2) & $>$ 99.8  (12.5)  & 75.9 (24.3) & 67.7 (23.1) \\
    & \multicolumn{1}{c|}{1}    & $>$ 99.8 (7.10)   & 83.3 (16.2) & $>$ 99.8 (7.70)    & 71.4 (17.5) & 69.2 (24.6) \\
 & \multicolumn{1}{c|}{2}    & $>$ 99.8 (0)   & $>$ 99.8 (13.3) &    $>$ 99.8 (0)   & 69.6 (16.4) & 69.2 (19.0) \\ \hline
 & \multicolumn{1}{c|}{0.25} & 1.70 (3.50) & $<$ 0.10 (0.80) & $<$ 0.10 (1.10)    &      $<$ 0.10 (0)  & $<$ 0.10 (0)       \\
 & \multicolumn{1}{c|}{0.5}  & 1.10 (2.00) & $<$ 0.10 (1.10)  & 3.50 (4.20) & $<$ 0.10 (0)         & $<$ 0.10 (0)         \\
FPR & \multicolumn{1}{c|}{0.75} & 1.10 (2.30) & $<$ 0.10 (1.10)    & 6.10 (7.70) & $<$ 0.10 (0)         & $<$ 0.10 (1.10)    \\
 & \multicolumn{1}{c|}{1}    & 1.20 (2.30) & $<$ 0.10 (16.2) & 8.10 (7.40) & $<$ 0.10 (0)         & $<$ 0.10 (16.2)      \\
 & \multicolumn{1}{c|}{2}    & 1.10 (2.30) &  $<$ 0.10 (1.10) & 11.6 (9.20) & $<$ 0.10 (0)         & $<$ 0.10 (0)      \\ \hline
\end{tabular}
\end{table}

\begin{table}[H]
\centering
\small
\begin{tabular}{|ccccccc|}
\hline
 &                           &               &                             & \hspace{-4cm}$\rho = 0.7$     &               &               \\
 & $\delta$                     & CoMPAdRe      & PAdRe & GAMSEL        & mSSL          & Lasso         \\ \hline
    & \multicolumn{1}{c|}{0.25} & 39.2 (23.6) & 16.7 (18.1) & 21.4 (27.9) & 13.3 (12.7) & 14.4 (20.4) \\
    & \multicolumn{1}{c|}{0.5}  & 84.6 (15.6) & 66.7 (19.2) & 80.0 (27.7) & 57.1 (23.5) & 56.3 (24.4) \\
TPR & \multicolumn{1}{c|}{0.75} & 91.7 (12.7) & 81.8 (15.6) & 96.9 (10.0) & 75.0 (17.0) & 70.0 (21.4) \\
    & \multicolumn{1}{c|}{1}    &  $>$ 99.8 (11.8)  & 84.0 (15.6) & $>$ 99.8 (7.60)    & 69.6 (19.0) & 67.9 (23.6)\\
 & \multicolumn{1}{c|}{2}    & $>$ 99.8 (0)   & $>$ 99.8 (8.30) &  $>$ 99.8 (0)  & 73.0 (13.0) & 74.3 (15.0) \\ \hline
 & \multicolumn{1}{c|}{0.25} & 0.60 (2.00) & $<$ 0.10 (0) & $<$ 0.10 (1.10)   &      $<$ 0.10 (0)  & $<$ 0.10 (0)       \\
 & \multicolumn{1}{c|}{0.5}  & 1.10 (1.10) & $<$ 0.10 (1.20) & 3.20 (4.40) & $<$ 0.10 (0)         & $<$ 0.10 (0)         \\
FPR & \multicolumn{1}{c|}{0.75} & 1.10 (2.30) & $<$ 0.10 (1.10)    & 6.70 (7.70) & $<$ 0.10 (0)         & $<$ 0.10 (1.10)    \\
 & \multicolumn{1}{c|}{1}    & $<$ 0.10 (1.10) & $<$ 0.10 (1.10) & 10.0 (8.40) & $<$ 0.10 (0)  & $<$ 0.10 (0.80)     \\
 & \multicolumn{1}{c|}{2}    & 1.10 (1.20) &  1.10 (1.20) & 11.0 (9.50) & $<$ 0.10 (0)         & $<$ 0.10 (0)      \\ \hline
\end{tabular}
\end{table}

\begin{table}[H]
\centering
\small
\begin{tabular}{|ccccccc|}
\hline
 &                           &               &                             & \hspace{-4cm}$\rho = 0.5$     &               &               \\
 & $\delta$                     & CoMPAdRe      & PAdRe & GAMSEL        & mSSL          & Lasso         \\ \hline
    & \multicolumn{1}{c|}{0.25} & 27.3 (27.3) & 17.9 (21.5) & 25.0 (34.0) & 12.9 (15.9) & 16.2 (18.5) \\
    & \multicolumn{1}{c|}{0.5}  & 77.8 (16.7) & 61.5 (21.4) & 77.8 (14.9) & 50.0 (24.1) & 54.7 (24.8) \\
TPR & \multicolumn{1}{c|}{0.75} & 87.5 (19.7) & 81.5 (23.7) & $>$ 99.8 (8.90) & 63.1 (18.0) & 63.1 (21.5) \\
    & \multicolumn{1}{c|}{1}    &  90.9 (14.3)  & 81.8 (17.3) & $>$ 99.8 (0)   & 68.8 (16.7) & 68.8 (19.3)\\
 & \multicolumn{1}{c|}{2}    &  $>$ 99.8 (0)  & 97.2 (10.0) &  $>$ 99.8 (0)  & 70.6 (24.0) & 69.6 (28.1) \\ \hline
 & \multicolumn{1}{c|}{0.25} & $<$ 0.10 (0) & $<$ 0.10 (0) & $<$ 0.10 (1.20)  & $<$ 0.10 (0)  & $<$ 0.10 (0)      \\
 & \multicolumn{1}{c|}{0.5}  & $<$ 0.10 (1.10) & $<$ 0.10 (1.20) & 3.40 (4.60) & $<$ 0.10 (0)         & $<$ 0.10 (0)         \\
FPR & \multicolumn{1}{c|}{0.75} & $<$ 0.10 (1.20) & $<$ 0.10 (1.10)   & 8.90 (8.40) & $<$ 0.10 (0)      & $<$ 0.10 (0.80)    \\
 & \multicolumn{1}{c|}{1}    & 1.10 (2.30) & 1.10 (1.20) & 10.1 (6.90) & $<$ 0.10 (0)  & $<$ 0.10 (1.10)     \\
 & \multicolumn{1}{c|}{2}    & 1.10 (2.30) &  $<$ 0.10 (1.20) & 13.5 (8.20) & $<$ 0.10 (0)         & $<$ 0.10 (0.80)    \\ \hline
\end{tabular}
\end{table}

\begin{table}[H]
\centering
\small
\begin{tabular}{|ccccccc|}
\hline
 &                           &               &                             & \hspace{-4cm}$\rho = 0.2$     &               &               \\
 & $\delta$                     & CoMPAdRe      & PAdRe & GAMSEL        & mSSL          & Lasso         \\ \hline
    & \multicolumn{1}{c|}{0.25} & 20.7 (25.0) & 19.1 (20.0) & 28.6 (32.0) & 11.1 (18.2) & 14.3 (18.5)\\
    & \multicolumn{1}{c|}{0.5}  & 64.0 (24.8) & 63.1 (23.1) & 76.0 (23.3) & 36.4 (16.4) & 50.0 (22.2) \\
TPR & \multicolumn{1}{c|}{0.75} & 83.8 (15.0) & 78.2 (14.4) & 96.9 (8.30) & 62.0 (27.1) & 64.3 (22.2)\\
    & \multicolumn{1}{c|}{1}    &  93.3 (15.1)  & 85.7 (21.4) & $>$ 99.8 (0)  & 65.6 (25.3) & 66.7 (26.5)\\
 & \multicolumn{1}{c|}{2}    &  $>$ 99.8 (6.60) & $>$ 99.8 (12.2) &  $>$ 99.8 (0)  & 69.6 (14.8) & 69.0 (15.4) \\ \hline
 & \multicolumn{1}{c|}{0.25} & $<$ 0.10 (1.10) & $<$ 0.10 (0) & 1.10 (2.30)  & $<$ 0.10 (0)  & $<$ 0.10 (0)      \\
 & \multicolumn{1}{c|}{0.5}  & $<$ 0.10 (1.10) & $<$ 0.10 (1.20) & 3.50 (3.60) & $<$ 0.10 (0)         & $<$ 0.10 (0)         \\
FPR & \multicolumn{1}{c|}{0.75} & $<$ 0.10 (1.20) & $<$ 0.10 (1.20)  & 7.00 (5.50) & $<$ 0.100 (0)      & $<$ 0.100 (1.10)   \\
 & \multicolumn{1}{c|}{1}    & $<$ 0.10 (1.20) & $<$ 0.10 (1.10) & 10.5 (8.00) & $<$ 0.10 (0)  & $<$ 0.10 (0)     \\
 & \multicolumn{1}{c|}{2}    & 1.10 (2.30) &  $<$ 0.10 (1.20) & 12.6 (6.10) & $<$ 0.10 (0)         & $<$ 0.10 (0)   \\ \hline
\end{tabular}
\caption{Summary of simulation results for settings where number of covariates $p = 10$. Results are divided into true positive rate (TPR) and false positive rate (FPR), expressed as a percentage, for levels of residual dependence $\rho = (0.2, 0.5, 0.7, 0.9)$ and signal-to-noise ratio $\delta = (0.25, 0.5, 0.75, 1, 2)$. Signal is divided into null vs. non-null signal. Results are presented as the median with the interquartile range in parenthesis. 50 datasets were simulated for each setting and sample size was fixed to $n = 250$ and the number of responses to $q = 10$.}
\end{table}

\newpage
\hspace{-0.75cm} \small \textbf{Estimation Accuracy:}
\begin{table}[H]
\centering
\small
\begin{tabular}{|ccccccc|}
\hline
 &                           &         $\rho = 0.9$              &           $\rho = 0.7$                     & $\rho = 0.5$     &     $\rho = 0.2$          &               \\ 
 & $\delta$                     & $\hat{f}$     & $\hat{f}$    & $\hat{f}$         & $\hat{f}$    &  \\ \hline
    & \multicolumn{1}{c|}{0.25} & 0.480 (0.195) & 0.858 (0.170) & 0.956 (0.161) & 0.999 (0.018) &  \\ 
   & \multicolumn{1}{c|}{0.5}  & 0.462 (0.179) & 0.589 (0.203) & 0.777 (0.169) & 0.989 (0.208) &\\
 & \multicolumn{1}{c|}{0.75} & 0.535 (0.248) & 0.750 (0.193) & 0.858 (0.160) & 0.960 (0.175) & \\
    & \multicolumn{1}{c|}{1}    &  0.546 (0.252)  & 0.700 (0.232) & 0.856 (0.261)  & 0.966 (0.167) & \\
 & \multicolumn{1}{c|}{2}    & 0.624 (0.221)  & 0.805 (0.216) &  0.858 (0.201) & 0.979 (0.159) & \\ \hline
\end{tabular}
\caption{\footnotesize Estimation accuracy results for non-linear signal $\Hat{f}$. Results represent the median ratio of mean absolute deviation (MAD):  ($\footnotesize{\frac{\hbox{CoMPAdRe}}{\hbox{PAdRe}}}$) across 50 simulated datasets per setting and the interquartile range is given in parenthesis. Settings considered vary based on signal-to-noise ratio $\delta$ and level of residual dependence $\rho$.}
\end{table}

\begin{table}[H]
\centering
\small
\begin{tabular}{|ccccccc|}
\hline
 &                           &         $\rho = 0.9$              &           $\rho = 0.7$                     & $\rho = 0.5$     &     $\rho = 0.2$          &               \\ 
 & $\delta$                     & $\hat{f}$     & $\hat{f}$    & $\hat{f}$         & $\hat{f}$    &  \\ \hline
    & \multicolumn{1}{c|}{0.25} & 0.380 (0.171) & 0.599 (0.223) & 0.732 (0.166) & 0.747 (0.101) &  \\ 
   & \multicolumn{1}{c|}{0.5}  & 0.355 (0.166) & 0.447 (0.175) & 0.539 (0.180) & 0.688 (0.222) &\\
 & \multicolumn{1}{c|}{0.75} & 0.347 (0.197) & 0.423 (0.213) & 0.464 (0.145) & 0.499 (0.154) & \\
    & \multicolumn{1}{c|}{1}    &  0.303 (0.147)  & 0.408 (0.141) & 0.480 (0.224)  & 0.485 (0.188) & \\
 & \multicolumn{1}{c|}{2}    &  0.333 (0.191) & 0.389 (0.277) &  0.454 (0.170) & 0.477 (0.181) & \\ \hline
\end{tabular}
\caption{\footnotesize Estimation accuracy results for non-linear signal $\Hat{f}$. Results represent the median ratio of mean absolute deviation (MAD):  ($\footnotesize{\frac{\hbox{CoMPAdRe}}{\hbox{GAMSEL}}}$) across 50 simulated datasets per setting with the interquartile range in parenthesis. Settings considered vary based on signal-to-noise ratio $\delta$ and level of residual dependence $\rho$.}
\end{table}

\begin{table}[H]
\centering
\small
\begin{tabular}{|ccccccc|}
\hline
 &                           &         $\rho = 0.9$              &           $\rho = 0.7$                     & $\rho = 0.5$     &     $\rho = 0.2$          &               \\ 
 & $\delta$                     & $\hat{f}$     & $\hat{f}$    & $\hat{f}$         & $\hat{f}$    &  \\ \hline
    & \multicolumn{1}{c|}{0.25} & 0.454 (0.153) & 0.802 (0.212) & 0.898 (0.110) & 0.935 (0.101) &  \\ 
   & \multicolumn{1}{c|}{0.5}  & 0.307 (0.157) & 0.408 (0.164) & 0.493 (0.129) & 0.598 (0.161) &\\
 & \multicolumn{1}{c|}{0.75} & 0.255 (0.130) & 0.307 (0.176) & 0.370 (0.118) & 0.401 (0.182) & \\
    & \multicolumn{1}{c|}{1}    &  0.202 (0.135) & 0.248 (0.115) &  0.298 (0.126) & 0.308 (0.118) & \\
 & \multicolumn{1}{c|}{2}    &  0.179 (0.089) & 0.208 (0.131) &  0.210 (0.070) & 0.221 (0.077) & \\ \hline
\end{tabular}
\caption{\footnotesize Estimation accuracy results for non-linear signal $\Hat{f}$. Results represent the median ratio of mean absolute deviation (MAD):  ($\footnotesize{\frac{\hbox{CoMPAdRe}}{\hbox{Lasso}}}$) across 50 simulated datasets per setting with the interquartile range in parenthesis. Settings considered vary based on signal-to-noise ratio $\delta$ and level of residual dependence $\rho$.}
\end{table}

\begin{table}[H]
\centering
\small
\begin{tabular}{|ccccccc|}
\hline
 &                           &         $\rho = 0.9$              &           $\rho = 0.7$                     & $\rho = 0.5$     &     $\rho = 0.2$          &               \\ 
 & $\delta$                     & $\hat{f}$     & $\hat{f}$    & $\hat{f}$         & $\hat{f}$    &  \\ \hline
    & \multicolumn{1}{c|}{0.25} & 0.613 (0.320) & 0.955 (0.104) & 1.00 (0.020) & 1.00 (0.035) &  \\ 
   & \multicolumn{1}{c|}{0.5}  & 0.415 (0.241) & 0.614 (0.278) & 0.596 (0.172) & 0.749 (0.204) &\\
 & \multicolumn{1}{c|}{0.75} & 0.350 (0.274) & 0.420 (0.178) & 0.447 (0.260) & 0.542 (0.302) & \\
    & \multicolumn{1}{c|}{1}    &  0.247 (0.163) & 0.342 (0.240) &  0.413 (0.211) & 0.454 (0.197) & \\
 & \multicolumn{1}{c|}{2}    &  0.233 (0.166) & 0.241 (0.159) &  0.248 (0.183) & 0.279 (0.215) & \\ \hline
\end{tabular}
\caption{\footnotesize Estimation accuracy results for non-linear signal $\Hat{f}$. Results represent the median ratio of mean absolute deviation (MAD):  ($\footnotesize{\frac{\hbox{CoMPAdRe}}{\hbox{mSSL}}}$) across 50 simulated datasets per setting with the interquartile range in parenthesis.. Settings considered vary based on signal-to-noise ratio $\delta$ and level of residual dependence $\rho$.}
\end{table}

\hspace{-0.75cm} \large \textbf{Linear Selection Accuracy:}

\begin{table}[H]
\centering
\small
\begin{tabular}{|cccccc|}
\hline
 &                           &             &             $\rho = 0.9$                &    &                    \\
 & $\delta$                     & CoMPAdRe      & PAdRe & GAMSEL    &        \\ \hline
    & \multicolumn{1}{c|}{0.25} & $>$ 99.7 (25.0) & $<$ 0.30 (13.8) & 6.30 (28.6) & \\
    & \multicolumn{1}{c|}{0.5}  & $>$ 99.7 (0) & 69.0 (50.0) & $>$ 99.7 (20.0) & \\
TPR & \multicolumn{1}{c|}{0.75} & $>$ 99.7 (0) & $>$ 99.7 (12.5) & $>$ 99.7 (0) & \\
    & \multicolumn{1}{c|}{1}    &  $>$ 99.7 (0) & $>$ 99.7 (12.5) & $>$ 99.7 (0) & \\
 & \multicolumn{1}{c|}{2}    &  $>$ 99.7 (0) & $>$ 99.7 (0) &  $>$ 99.7 (0) & \\ \hline
 & \multicolumn{1}{c|}{0.25} & 0.50 (1.20) & $<$ 0.1 (0) & $<$ 0.10 (0.10)  &  \\
 & \multicolumn{1}{c|}{0.5}  & $<$ 0.10  (1.10) & $<$ 0.10  (0) & 2.30 (3.40) &   \\
FPR & \multicolumn{1}{c|}{0.75} & $<$ 0.10  (1.90) & $<$ 0.10  (1.10)  & 3.50 (5.70) &  \\
 & \multicolumn{1}{c|}{1}    & $<$ 0.10  (1.20) & $<$ 0.10  (0) & 6.30 (4.70) &  \\
 & \multicolumn{1}{c|}{2}    & $<$ 0.10  (1.20) & $<$ 0.10  (0) & 7.90 (4.70) & \\ \hline
\end{tabular}
\end{table}

\begin{table}[H]
\centering
\small
\begin{tabular}{|cccccc|}
\hline
 &                           &             &             $\rho = 0.7$                &    &                    \\
 & $\delta$                     & CoMPAdRe      & PAdRe & GAMSEL    &        \\ \hline
    & \multicolumn{1}{c|}{0.25} & 35.4 (35.7) & $<$ 0.30 (8.30)  & 6.30 (28.6) & \\
    & \multicolumn{1}{c|}{0.5}  & $>$ 99.7 (11.0) & 73.9 (31.1) & 90.0 (21.7) & \\
TPR & \multicolumn{1}{c|}{0.75} & $>$ 99.7 (0) & $>$ 99.7 (8.30) & $>$ 99.7 (0) & \\
    & \multicolumn{1}{c|}{1}    &  $>$ 99.7 (0) & $>$ 99.7 (12.2) & $>$ 99.7 (0) & \\
 & \multicolumn{1}{c|}{2}    & $>$ 99.7 (0) & $>$ 99.7 (0) &  $>$ 99.7 (0) & \\ \hline
 & \multicolumn{1}{c|}{0.25} & $<$ 0.10  (1.10) & $<$ 0.10  (0) & $<$ 0.10  (0.10)  &  \\
 & \multicolumn{1}{c|}{0.5}  & $<$ 0.10  (1.10) & $<$ 0.10  (0) & 1.20 (4.30) &   \\
FPR & \multicolumn{1}{c|}{0.75} & $<$ 0.10  (1.20) & $<$ 0.10  (0.80)  & 4.60 (5.70) &  \\
 & \multicolumn{1}{c|}{1}    & $<$ 0.10  (0.90) &  $<$ 0.10  (0.80) & 6.20 (5.00) &  \\
 & \multicolumn{1}{c|}{2}    & $<$ 0.10  (1.10) & $<$ 0.10  (0) & 7.90 (6.70) & \\ \hline
\end{tabular}
\end{table}

\begin{table}[H]
\centering
\small
\begin{tabular}{|cccccc|}
\hline
 &                           &             &             $\rho = 0.5$                &    &                    \\
 & $\delta$                     & CoMPAdRe      & PAdRe & GAMSEL    &        \\ \hline
    & \multicolumn{1}{c|}{0.25} & 11.1 (36.4) & $<$ 0.30 (14.3)  & 13.4 (28.8) & \\
    & \multicolumn{1}{c|}{0.5}  & 88.9 (16.7) & 71.4 (33.3) & $>$ 99.7 (16.7) & \\
TPR & \multicolumn{1}{c|}{0.75} & $>$ 99.7 (0) & $>$ 99.7 (16.7) & $>$ 99.7 (0) & \\
    & \multicolumn{1}{c|}{1}    &  $>$ 99.7 (0) & $>$ 99.7 (0) & $>$ 99.7 (0) & \\
 & \multicolumn{1}{c|}{2}    & $>$ 99.7 (0) & $>$ 99.7 (0) &  $>$ 99.7 (0) & \\ \hline
 & \multicolumn{1}{c|}{0.25} & $<$ 0.10 (0) & $<$ 0.10 (0) & $<$ 0.10 (0)  &  \\
 & \multicolumn{1}{c|}{0.5}  & $<$ 0.10 (0) & $<$ 0.10 (0) & 2.30 (4.22) &   \\
FPR & \multicolumn{1}{c|}{0.75} & $<$ 0.10 (1.10) & $<$ 0.10 (0.80)  & 6.60 (6.40) &  \\
 & \multicolumn{1}{c|}{1}    & $<$ 0.10 (1.20) &  $<$ 0.10 (1.10) & 7.90 (5.30) &  \\
 & \multicolumn{1}{c|}{2}    & $<$ 0.10 (1.10) & $<$ 0.10 (0.80) & 8.20 (5.50) & \\ \hline
\end{tabular}
\end{table}

\begin{table}[H]
\centering
\small
\begin{tabular}{|cccccc|}
\hline
 &                           &             &             $\rho = 0.2$                &    &                    \\
 & $\delta$                     & CoMPAdRe      & PAdRe & GAMSEL    &        \\ \hline
    & \multicolumn{1}{c|}{0.25} & $<$ 0.30 (25.0) & $<$ 0.30 (20.0)  & 20.0 (40.0) & \\
    & \multicolumn{1}{c|}{0.5}  & 71.4 (42.5) & 70.7 (26.8) & 87.5 (25.9) & \\
TPR & \multicolumn{1}{c|}{0.75} & $>$ 99.7 (0) & $>$ 99.7 (20.0) & $>$ 99.7 (0) & \\
    & \multicolumn{1}{c|}{1}    & $>$ 99.7 (0) & $>$ 99.7 (0) & $>$ 99.7 (0) & \\
 & \multicolumn{1}{c|}{2}    & $>$ 99.7 (0) & $>$ 99.7 (0) &  $>$ 99.7 (0) & \\ \hline
 & \multicolumn{1}{c|}{0.25} & $<$ 0.10  (0) & $<$ 0.10  (20.0) & $<$ 0.10  (1.20)  &  \\
 & \multicolumn{1}{c|}{0.5}  & $<$ 0.10  (1.10) & $<$ 0.10  (0)& 3.40 (3.60) &   \\
FPR & \multicolumn{1}{c|}{0.75} & $<$ 0.10  (1.10) & $<$ 0.10  (1.10) & 5.80 (5.40) &  \\
 & \multicolumn{1}{c|}{1}    & $<$ 0.10  (0) &  $<$ 0.10  (0) & 6.90 (5.50) &  \\
 & \multicolumn{1}{c|}{2}    & $<$ 0.10  (1.10) & $<$ 0.10  (0) & 8.20 (5.30) & \\ \hline
\end{tabular}
\caption{\textbf{Linear Selection Results: }Summary of selection results for Linear functions for methods that select both Linear and Non-Linear functions in settings where number of covariates $p = 10$. Results are divided into true positive rate (TPR) and false positive rate (FPR), expressed as a percentage, for levels of residual dependence $\rho = (0.2, 0.5, 0.7, 0.9)$ and signal-to-noise ratio $\delta = (0.25, 0.5, 0.75, 1, 2)$. Results are presented as the median with the interquartile range
in parenthesis. 50 datasets were simulated for each setting and sample size was fixed to $n = 250$ and the number of responses to $q = 10$.}
\end{table}

\newpage
\hspace{-0.75cm} \large \textbf{Non-Linear Selection Accuracy:}

\begin{table}[H]
\centering
\small
\begin{tabular}{|cccccc|}
\hline
 &                           &             &             $\rho = 0.9$                &    &                    \\
 & $\delta$                     & CoMPAdRe      & PAdRe & GAMSEL    &        \\ \hline
    & \multicolumn{1}{c|}{0.25} & 50.0 (26.7) & $<$ 0.30 (14.3)  & $<$ 0.30 (26.7) & \\
    & \multicolumn{1}{c|}{0.5}  & 66.7 (28.3) & 46.4 (34.1) & 57.1 (41.8) & \\
TPR & \multicolumn{1}{c|}{0.75} & 75.0 (27.8) & 57.1 (27.4) & 73.2 (37.5) & \\
    & \multicolumn{1}{c|}{1}    &  72.8 (19.1) & 52.3 (26.7) & 82.6 (23.7) & \\
 & \multicolumn{1}{c|}{2}    & 86.6 (25.0) & 71.4 (27.3) &  $>$ 99.7 (25.0) & \\ \hline
 & \multicolumn{1}{c|}{0.25} & 1.00 (2.00) & $<$ 0.10 (0) & $<$ 0.10 (0)  &  \\
 & \multicolumn{1}{c|}{0.5}  & $<$ 0.10 (1.10) & $<$ 0.10 (0) & $<$ 0.10 (2.10) &   \\
FPR & \multicolumn{1}{c|}{0.75} & $<$ 0.10 (1.10) & $<$ 0.10 (0)  & 2.80 (7.20) &  \\
 & \multicolumn{1}{c|}{1}    & $<$ 0.10 (1.10) &  $<$ 0.10 (0) & 4.30 (7.50) &  \\
 & \multicolumn{1}{c|}{2}    & $<$ 0.10 (1.10) & $<$ 0.10 (1.10) & 6.30 (7.50) & \\ \hline
\end{tabular}
\end{table}

\begin{table}[H]
\centering
\small
\begin{tabular}{|cccccc|}
\hline
 &                           &             &             $\rho = 0.7$                &    &                    \\
 & $\delta$                     & CoMPAdRe      & PAdRe & GAMSEL    &        \\ \hline
    & \multicolumn{1}{c|}{0.25} & 25.0 (37.5) & $<$ 0.30 (12.5) & $<$ 0.30 (11.9) & \\
    & \multicolumn{1}{c|}{0.5}  & 50.0 (33.3) & 33.3 (32.5) & 40.0 (39.7) & \\
TPR & \multicolumn{1}{c|}{0.75} & 60.0 (25.0) & 50.0 (29.2) & 66.7 (30.0) & \\
    & \multicolumn{1}{c|}{1}    &  66.7 (29.5) & 60.0 (25.0) & 78.9 (31.3) & \\
 & \multicolumn{1}{c|}{2}    & 87.5 (20.0) & 80.0 (21.5) &  $>$ 99.7 (16.7) & \\ \hline
 & \multicolumn{1}{c|}{0.25} & $<$ 0.10 (0) & $<$ 0.10 (0) & $<$ 0.10 (0)  &  \\
 & \multicolumn{1}{c|}{0.5}  & $<$ 0.10 (1.10) & $<$ 0.10 (0) & 0.50 (2.20) &   \\
FPR & \multicolumn{1}{c|}{0.75} & $<$ 0.10 (1.10) & $<$ 0.10 (0)  & 2.20 (5.40)&  \\
 & \multicolumn{1}{c|}{1}    & $<$ 0.10 (1.00) &  $<$ 0.10 (0) & 6.40 (7.60) &  \\
 & \multicolumn{1}{c|}{2}    & $<$ 0.10 (1.10) & $<$ 0.10 (1.10) & 6.90 (5.30) & \\ \hline
\end{tabular}
\end{table}

\begin{table}[H]
\centering
\small
\begin{tabular}{|cccccc|}
\hline
 &                           &             &             $\rho = 0.5$                &    &                    \\
 & $\delta$                     & CoMPAdRe      & PAdRe & GAMSEL    &        \\ \hline
    & \multicolumn{1}{c|}{0.25} & 5.00 (20.0) & $<$ 0.30 (12.5) & $<$ 0.30 (16.7) & \\
    & \multicolumn{1}{c|}{0.5}  & 50.0 (33.3) & 37.5 (28.2) & 43.7 (23.8) & \\
TPR & \multicolumn{1}{c|}{0.75} & 60.0 (21.4) & 57.3 (16.7) & 70.7 (35.7) & \\
    & \multicolumn{1}{c|}{1}    &  62.5 (30.6) & 50.0 (31.4) & 80.0 (33.3) & \\
 & \multicolumn{1}{c|}{2}    & 83.3 (28.6) & 75.0 (32.5) &  $>$ 99.7 (17.8) & \\ \hline
 & \multicolumn{1}{c|}{0.25} &  $<$ 0.10 (0) &  $<$ 0.10 (0) &  $<$ 0.10 (0.008) &  \\
 & \multicolumn{1}{c|}{0.5}  &  $<$ 0.10 (0) &  $<$ 0.10 (0) & 1.00 (2.10) &   \\
FPR & \multicolumn{1}{c|}{0.75} &  $<$ 0.10 (0.80) &  $<$ 0.10 (0)  & 3.80 (4.90) &  \\
 & \multicolumn{1}{c|}{1}    &  $<$ 0.10 (1.10) &  $<$ 0.10 (1.10) & 5.50 (5.30) &  \\
 & \multicolumn{1}{c|}{2}    &  $<$ 0.10 (1.00) &  $<$ 0.10 (1.00) & 10.1 (10.2) & \\ \hline
\end{tabular}
\end{table}

\begin{table}[H]
\centering
\small
\begin{tabular}{|cccccc|}
\hline
 &                           &             &             $\rho = 0.2$                &    &                    \\
 & $\delta$                     & CoMPAdRe      & PAdRe & GAMSEL    &        \\ \hline
    & \multicolumn{1}{c|}{0.25} & $<$ 0.30 (10.0) & $<$ 0.30 (10.0) & $<$ 0.30 (16.1) & \\
    & \multicolumn{1}{c|}{0.5}  & 40.0 (43.8) & 40.0 (36.7) & 50.0 (30.2) & \\
TPR & \multicolumn{1}{c|}{0.75} & 50.0 (26.7) & 50.0 (25.6) & 66.7 (30.0) & \\
    & \multicolumn{1}{c|}{1}    & 62.5 (21.4) & 62.5 (24.1) & 80.0 (21.9) & \\
 & \multicolumn{1}{c|}{2}    & 75.0 (39.4) & 75.0 (27.1) & 96.2 (20.0) & \\ \hline
 & \multicolumn{1}{c|}{0.25} & $<$ 0.10 (0) & $<$ 0.10 (0) & $<$ 0.10 (1.10) &  \\
 & \multicolumn{1}{c|}{0.5}  & $<$ 0.10 (0) & $<$ 0.10 (0.30) & 1.10 (2.20) &   \\
FPR & \multicolumn{1}{c|}{0.75} &  $<$ 0.10 (0) & $<$ 0.10 (0)  & 2.10 (4.30) &  \\
 & \multicolumn{1}{c|}{1}    & $<$ 0.10 (0) & $<$ 0.10 (0) & 7.30 (6.70) &  \\
 & \multicolumn{1}{c|}{2}    & $<$ 0.10 (1.10) & $<$ 0.10 (1.10) & 8.70 (8.60) & \\ \hline
\end{tabular}
\caption{\textbf{Non-Linear Selection Results: }Summary of selection results for Non-Linear functions for methods that select both Linear and Non-Linear functions in settings where number of covariates $p = 10$. Results are divided into true positive rate (TPR) and false positive rate (FPR), expressed as a percentage, for levels of residual dependence $\rho = (0.2, 0.5, 0.7, 0.9)$ and signal-to-noise ratio $\delta = (0.25, 0.5, 0.75, 1, 2)$. Results are presented as the median with the interquartile range
in parenthesis. 50 datasets were simulated for each setting and sample size was fixed to $n = 250$ and the number of responses to $q = 10$.}
\end{table}

\noindent \textbf{Tuning Parameter Selection:}  All tuning parameters for selection in CoMPAdRe were chosen via cross-validation. To avoid overfitting and to encourage parsimonious model fits, hyperparameter values were selected to produce the sparsest model within 1 standard deviation of the value that minimized mean squared error as commonly recommended in practice. We also note that CoMPAdRe, for higher dimensional predictor settings ($p = 100$), produced starting values where the range of hyperparameters considered for initial non-linear selection was no more than a factor of 0.75 smaller than the smallest value to produce a null solution (i.e. no non-linear predictor associations initially selected). This choice of hyper-parameter range when initializing the algorithm helped encourage numerical stability in high dimensional settings (a common challenge encountered in multivariate selection approaches as dimension is increased). We ran CoMPAdRe for 5 iterations for all simulations considered.  

\subsubsection*{\large S.1.2: Additional Function-Specific Simulations}
We design a simple simulation to assess the performance of CoMPAdRe in selecting and estimating specific functional shapes relative to competitors. Figure 6 visualizes the shapes considered, the same as those in the main body of the text. We let Y be a $n \times q$ matrix of $q$ responses and $X$ a  $n \times p$ matrix of $p$ predictors and set $q = 10$ and $p = 10$. In this simulation, we only set two covariate response combinations, $(Y[,1] - X[,1], Y[,2] - X[,2])$, to be non-sparse with identical true signal (for both magnitude and functional shape): $Y_{m} = \delta * f_{j}$ for $m = (1, 2)$, $j = (1, 2, 3, 4, 5)$, and $\delta = (0.5, 2)$. We simulate $X$ and $E$ as in the main body of this text, fixing $\rho = 0.7$ to induce moderate dependence among responses. We simulate 50 datasets per setting considered and run CoMPAdRe for five iterations. 

Results demonstrate that, in this simplified example, CoMPAdRe generally outperformed competitors in terms of both estimation and selection accuracy for each functional shape considered. In particular, linear selection methods (Lasso, mSSL) failed to select functions $f_{2}$ and $f_{4}$, even at high levels of signal-to-noise $\delta = 2$. Furthermore, these linear selection approaches showed worse estimation accuracy than CoMPAdRe, with differences most stark for $f_{1}$, $f_{2}$, and $f_{4}$. We note that GAMSEL does not re-estimate post selection as CoMPAdRe, PAdRe, mSSL, and the Lasso (where we re-estimated post-selection as often recommended in practice). This difference likely negatively contributes to the estimation accuracy of GAMSEL for estimating certain functions $f_{j}$. We also note that these simulation studies are highly sparse and simplified. CoMPAdRe performed even better relative to competitors in the simulations conducted in the main body of this manuscript, suggesting the method's performance may be even better compared to other approaches in more realistic, complex settings with non-trivial residual dependence among responses. 
\vspace{1cm}

\begin{figure}[H]
\centerline{\includegraphics[height=8cm,width=11cm]{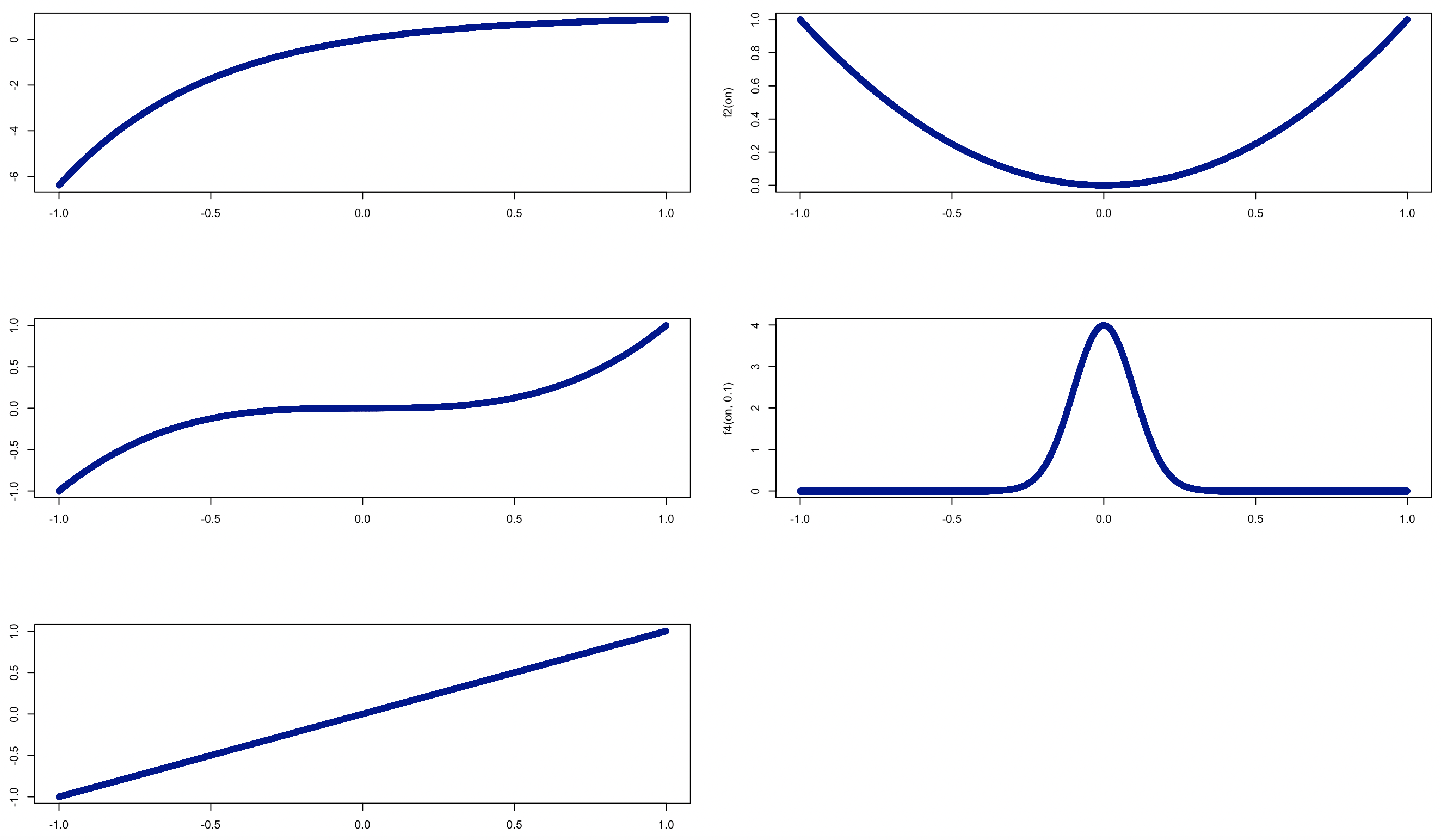}}
\caption{The shapes of covariate-response associations considered in our simulation study.}\label{fig:01}
\end{figure}

\newpage
\textbf{Selection Performance} 
\begin{table}[H]
\small
\setstretch{1.5}
\begin{center}
\begin{tabular}{|cccccc|}
\hline
              &          &   $\boldsymbol{f_{1}}=\delta*(1-\hbox{exp}(-2X))$   &      &        &       \\
$\delta$                     & CoMPAdRe & PAdRe & mSSL  & GAMSEL & Lasso \\ \hline
\multicolumn{1}{|c|}{0.5} & $>$ 99.0 (0)    & $>$ 99.0 (0) & $>$ 99.0 (0) & $>$ 99.0 (0)  & $>$ 99.0 (0) \\
\multicolumn{1}{|c|}{2}   & $>$ 99.0 (0)    & $>$ 99.0 (0) & $>$ 99.0 (0) & $>$ 99.0 (0)  & $>$ 99.0 (0) \\ \hline
\end{tabular}
\end{center}
\end{table}

\begin{table}[H]
\setstretch{1.5}
\begin{center}
\begin{tabular}{|cccccc|}
\hline
           &          &   $\boldsymbol{f_{2}}=\delta*(X^{2})$    &       &        &       \\
$\delta$                     & CoMPAdRe & PAdRe & mSSL  & GAMSEL & Lasso \\ \hline
\multicolumn{1}{|c|}{0.5} & $<$ 1.00 (0)    & $<$ 1.00 (0) & $<$ 1.00 (0) &  $<$ 1.00 (0) & $<$ 1.00 (0) \\
\multicolumn{1}{|c|}{2}   & $>$ 99.0 (0)    & $>$ 99.0 (0) & $<$ 1.00 (0) & $>$ 99.0 (0)  & $<$ 1.00 (0) \\ \hline
\end{tabular}
\end{center}
\end{table}

\begin{table}[H]
\setstretch{1.5}
\begin{center}
\begin{tabular}{|cccccc|}
\hline
             &          &  $\boldsymbol{f_{3}}=\delta*(X^{3})$     &       &        &       \\
$\delta$                     & CoMPAdRe & PAdRe & mSSL  & GAMSEL & Lasso \\ \hline
\multicolumn{1}{|c|}{0.5} & 50.0 (50.0)    & $<$ 1.00 (0) & $<$ 1.00 (0) &  $<$ 1.00 (0) & $<$ 1.00 (0) \\
\multicolumn{1}{|c|}{2}   & $>$ 99.0 (0)    & $>$ 99.0 (0) & $<$ 1.00 (0) & $>$ 99.0 (0)  & $<$ 1.00 (0) \\ \hline
\end{tabular}
\end{center}
\end{table}

\begin{table}[H]
\setstretch{1.5}
\small
\begin{center}
\begin{tabular}{|cccccc|}
\hline
              &          &  $\boldsymbol{f_{4}}(\sigma = 0.1)= \delta*\frac{1}{\sqrt{2\pi}\sigma}\hbox{exp}(-\frac{X^{2}}{2\sigma^{2}}))$  &       &        &       \\
$\delta$                     & CoMPAdRe & PAdRe & mSSL  & GAMSEL & Lasso \\ \hline
\multicolumn{1}{|c|}{0.5} & $>$ 99.0 (0)    & $>$ 99.0 (0) & $<$ 1.00 (0) &  $>$ 99.0 (50.0) & $<$ 1.00 (0) \\
\multicolumn{1}{|c|}{2}   & $>$ 99.0 (0)    & $>$ 99.0 (0) & $<$ 1.00 (0) & $>$ 99.0 (0)  & $<$ 1.00 (0) \\ \hline
\end{tabular}
\end{center}
\end{table}

\begin{table}[H]
\setstretch{1.5}
\begin{center}
\begin{tabular}{|cccccc|}
\hline
            &          &   $\boldsymbol{f_{5}}=\delta*(X)$     &       &        &       \\
$\delta$                     & CoMPAdRe & PAdRe & mSSL  & GAMSEL & Lasso \\ \hline
\multicolumn{1}{|c|}{0.5} & $>$ 99.0 (0)    & 50.0 (50.0) & $>$ 99.0 (0) &  50.0 (50.0) & 50.0 (50.0) \\
\multicolumn{1}{|c|}{2}   & $>$ 99.0 (0)    & $>$ 99.0 (0) & $>$ 99.0 (0) & $>$ 99.0 (0)  & $>$ 99.0 (0) \\ \hline
\end{tabular}
\caption{True positive rates, expressed as a percentage, for settings described in supplemental section \textbf{S.1.2}.  50 datasets were simulated per setting and each setting has two non-sparse covariate-response associations with identical magnitude and functional form.  Values represent the median true positive rate across 50 simulated datasets, with the interquartile range (IQR) of observed true positive rates shown in parenthesis.}
\end{center}
\end{table}

\newpage

\textbf{Estimation Performance} 

\begin{table}[H]
\setstretch{1.5}
\begin{center}
\begin{tabular}{|cccccc|}
\hline
           &          &       &   $\boldsymbol{f_{1}}=\delta*(1-\hbox{exp}(-2X))$     &        &       \\
$\delta$                     & CoMPAdRe & PAdRe & mSSL  & GAMSEL & Lasso \\ \hline
\multicolumn{1}{|c|}{0.5} & 0.016 (0.005)    & 0.083 (0.005) & 0.083 (0.006) &  0.086 (0.005) & 0.083 (0.005) \\
\multicolumn{1}{|c|}{2}   & 0.019 (0.005)    & 0.027 (0.007) & 0.329 (0.021) & 0.093 (0.010)  & 0.330 (0.019) \\ \hline
\end{tabular}
\end{center}
\end{table}

\begin{table}[H]
\setstretch{1.5}
\begin{center}
\begin{tabular}{|cccccc|}
\hline
             &          &       &  $\boldsymbol{f_{2}}=\delta*(X^{2})$     &        &       \\
$\delta$                     & CoMPAdRe & PAdRe & mSSL  & GAMSEL & Lasso \\ \hline
\multicolumn{1}{|c|}{0.5} &  0.033 (0.002)   & 0.033 (0.002) & 0.033 (0.002) & 0.060 (0.018)  & 0.033 (0.002) \\
\multicolumn{1}{|c|}{2}   &  0.014 (0.006)  & 0.020 (0.010) & 0.133 (0.006) & 0.085 (0.028)  & 0.133 (0.006) \\ \hline
\end{tabular}
\end{center}
\end{table}

\begin{table}[H]
\setstretch{1.5}
\begin{center}
\begin{tabular}{|cccccc|}
\hline
          &          &      &    $\boldsymbol{f_{3}}=\delta*(X^{3})$      &        &       \\
$\delta$                     & CoMPAdRe & PAdRe & mSSL  & GAMSEL & Lasso \\ \hline
\multicolumn{1}{|c|}{0.5} &  0.023 (0.001)   & 0.025 (0.001) & 0.024 (0.002) & 0.069 (0.026)  & 0.025 (0.001) \\
\multicolumn{1}{|c|}{2}   &  0.021 (0.018)  & 0.046 (0.014) & 0.051 (0.004) & 0.083 (0.019)  & 0.052 (0.004) \\ \hline
\end{tabular}
\end{center}
\end{table}

\begin{table}[H]
\setstretch{1.5}
\begin{center}
\begin{tabular}{|cccccc|}
\hline
            &          &      &   $\boldsymbol{f_{4}}(\sigma = 0.1)=\delta*\frac{1}{\sqrt{2\pi}\sigma}\hbox{exp}(-\frac{X^{2}}{2\sigma^{2}}))$     &        &       \\
$\delta$                     & CoMPAdRe & PAdRe & mSSL  & GAMSEL & Lasso \\ \hline
\multicolumn{1}{|c|}{0.5} &  0.026 (0.006)   & 0.037 (0.007) & 0.049 (0.006) & 0.109 (0.024)  & 0.049 (0.006) \\
\multicolumn{1}{|c|}{2}   &  0.056 (0.017)  & 0.063 (0.021) & 0.201 (0.025) & 0.214 (0.022)  & 0.201 (0.025) \\ \hline
\end{tabular}
\end{center}
\end{table}

\begin{table}[H]
\setstretch{1.5}
\begin{center}
\begin{tabular}{|cccccc|}
\hline
           &          &    &    $\boldsymbol{f_{5}}=\delta*(X)$     &        &       \\
$\delta$                     & CoMPAdRe & PAdRe & mSSL  & GAMSEL & Lasso \\ \hline
\multicolumn{1}{|c|}{0.5} &  0.008 (0.011)   & 0.041 (0.021) & 0.005 (0.003) & 0.081 (0.018) & 0.032 (0.021) \\
\multicolumn{1}{|c|}{2}   &  0.005 (0.004) & 0.007 (0.007) & 0.005 (0.004) & 0.085 (0.017) & 0.007 (0.008) \\ \hline
\end{tabular}
\caption{Mean absolute deviation (MAD) for settings described in supplemental section \textbf{S.1.2}. 50 datasets were simulated per setting and each setting has two non-sparse covariate-response associations with identical magnitude and functional form.  Values represent the median mean absolute deviation (MAD) across 50 simulated datasets, with the interquartile range (IQR) of observed mean absolute deviation shown in parenthesis.}
\end{center}
\end{table}

\newpage
\subsection*{S.2: CoMPAdRe analysis of proteomics application}

We first visualize plots of non-linear mRNA-protein associations found across all 8 breast cancer pathways below.

\begin{figure}[H]
\centerline{\includegraphics[height=10cm,width=13cm]{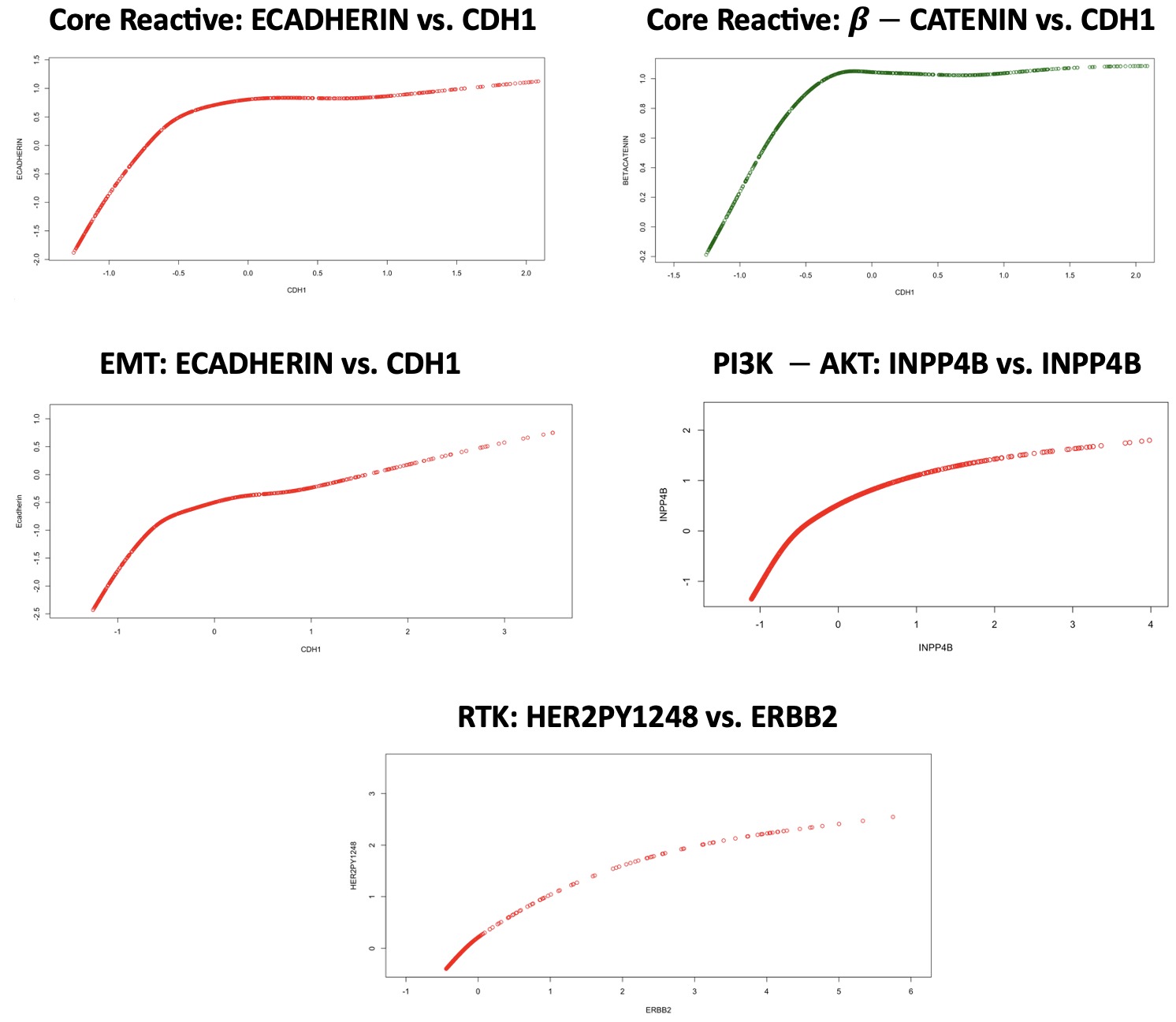}}
\caption{The shapes of non-linear protein-mRNA associations (Protein vs. mRNA) found across 8 breast cancer pathways from the cancer proteome atlas (TCPA).}\label{fig:01}
\end{figure}

\noindent We next show protein-protein covariance networks for the breast cancer pathways not shown in the main body of this manuscript. 

\begin{figure}[H]
\centerline{\includegraphics[height=15cm,width=17cm]{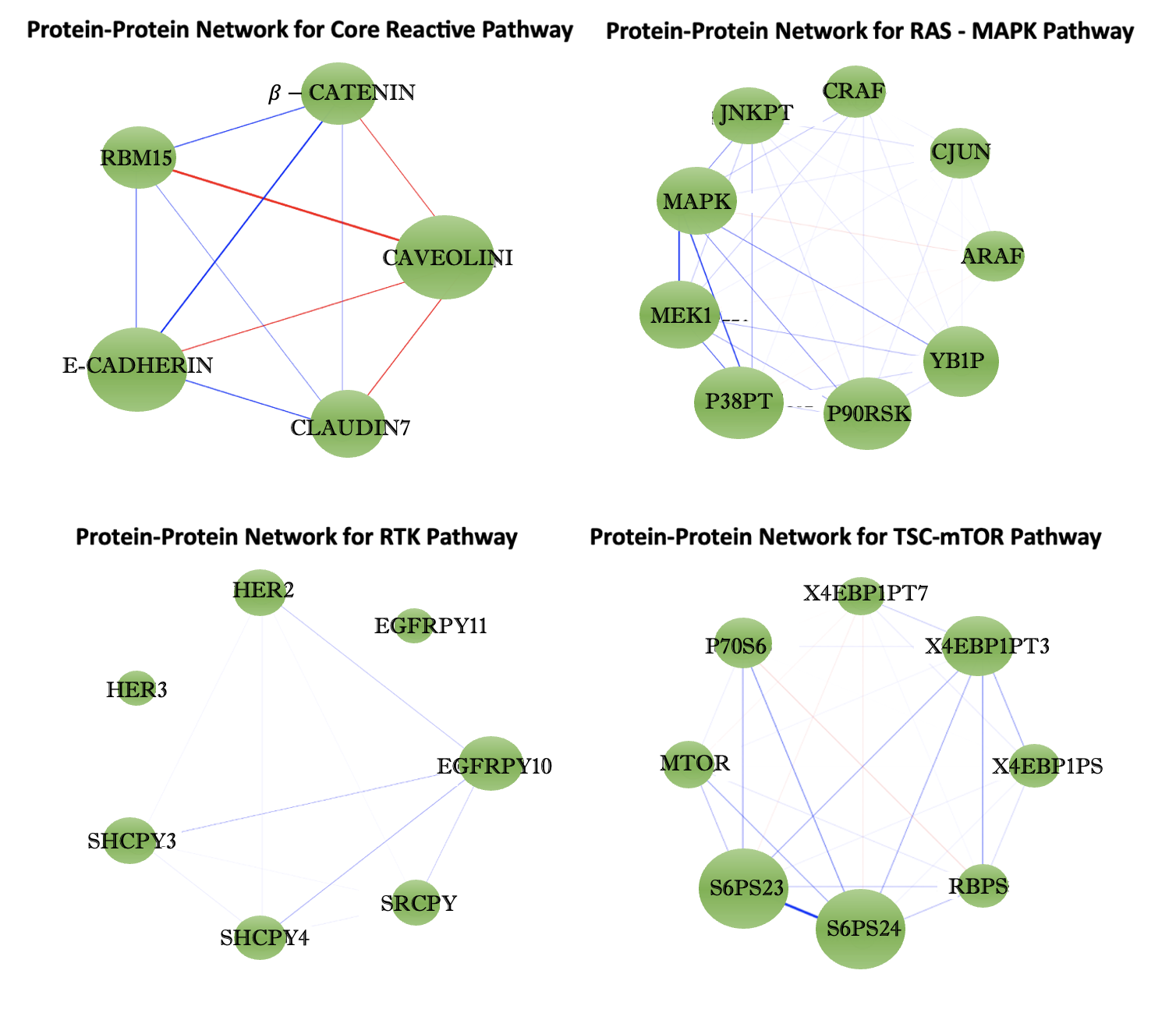}}
\caption{\setstretch{1.1}Protein-Protein covariance networks for Core Reactive, RAS-MAPK, RTK, and TSC-mTOR pathways. Blue edges indicate negative associations while red edges indicate positive associations. Edge thickness indicates the magnitude of the dependence between two corresponding proteins and node size is scaled relative to the strength and number of connections for a protein.}\label{fig:01}
\end{figure}

\noindent We conclude this section with a full comprehensive table of linear and non-linear Protein-mRNA associations found across all 8 pathways, as seen originally in the main body of this manuscript.
\begin{table}[H]
\tiny
\centering
\setstretch{1.7}
\begin{adjustbox}{angle=90}
\begin{tabular}{|c|c|c|}
\hline
pathway         & linear mRNA selected                               & non-linear mRNA selected         \\ \hline
Breast Reactive & GAPDH-GAPDH                                        & None                             \\
Core Reactive   & None                                               & CDH1-Betacatenin, CDH1-Ecadherin \\
DNA damage      & RAD50-RAD50, MRE11A-RAD50, ATM-ATM, TP53BP1-X53BP1 & None                             \\
EMT             & CDH1-Betacatenin                                   & CDH1-Ecadherin                   \\
PI3K - AKT & PTEN-PTEN, CDK1B-AKTPS473, CDK1B-AKTPT308, AKT1-AKTPS473, AKT1-AKTPT308, GSK3B-INPP4B, CDK1B-INPP4B & INPP4B-INPP4B \\
RAS-MAPK        & YBX1-JNKPT183Y185,  YBX1-YB1PS102                  & None                             \\
RTK             & ERBB2-EGFRPY1068, EGFR-EGFRPY1068                  & ERBB2-HER2PY1248                 \\
TSC-mTOR   & EIF4EBP1-X4EBP1PS65, EIF4EBP1-X4EBP1PT37T46, EIF4EBP1-X4EBP1PT70                                    & None          \\ \hline
\end{tabular}
\end{adjustbox}
\end{table}

\end{document}